\documentclass[prx, twocolumn, superscriptaddress,
 showpacs,
 longbibliography,
 aps,
 showkeys]{revtex4-2}

\usepackage{hyperref,color,graphicx}
\usepackage{amsfonts,amssymb,amsmath}
\usepackage[english]{babel}
\usepackage{xcolor}
\usepackage{mathrsfs}
\usepackage{graphicx,float}
\usepackage[space]{grffile}
\usepackage{placeins}
\usepackage{latexsym}
\usepackage{textcomp}
\usepackage{longtable}
\usepackage{multirow,booktabs}
\usepackage{url}
\hypersetup{colorlinks=true,linkcolor=blue,citecolor=blue}
\usepackage[english]{babel}
\usepackage{dcolumn}
\usepackage{bm}
\usepackage{units}
\usepackage{upgreek}
\usepackage{color}
\usepackage{cancel}
\usepackage{hyperref}
\usepackage{braket}

\usepackage{centernot}
\begin{document}

\title{Single-photon source over the terahertz regime}

\author{Caspar Groiseau}
\email{caspar.groiseau@uam.es}
\affiliation{Departamento de F\'isica Teórica de la Materia Condensada and Condensed Matter Physics Center (IFIMAC), Universidad Autónoma de Madrid, 28049 Madrid, Spain}

\author{Antonio I. Fernández-Domínguez}
\affiliation{Departamento de F\'isica Teórica de la Materia Condensada and Condensed Matter Physics Center (IFIMAC), Universidad Autónoma de Madrid, 28049 Madrid, Spain}

\author{Diego Martín-Cano}
\email{diego.martin.cano@uam.es}
\affiliation{Departamento de F\'isica Teórica de la Materia Condensada and Condensed Matter Physics Center (IFIMAC), Universidad Autónoma de Madrid, 28049 Madrid, Spain}

\author{Carlos Sánchez Muñoz}
\email{carlos.sanchezmunnoz@uam.es}
\affiliation{Departamento de F\'isica Teórica de la Materia Condensada and Condensed Matter Physics Center (IFIMAC), Universidad Autónoma de Madrid, 28049 Madrid, Spain}

\begin{abstract}
We present a proposal for a tunable source of single photons operating in the terahertz (THz) regime. 
This scheme transforms incident visible photons into quantum THz radiation by driving a single polar quantum emitter with an optical laser, with its permanent dipole enabling dressed THz transitions enhanced by the resonant coupling to a cavity. 
This mechanism offers optical tunability of properties such as the frequency of the emission or its quantum statistics (ranging from antibunching to entangled multi-photon states) by modifying the intensity and frequency of the drive. We show that the implementation of this proposal is feasible with state-of-the-art photonics technology. 

\end{abstract}

\keywords{THz radiation, Quantum optics, Nanophotonics, Polar emitter, Correlated photons}

\maketitle

\emph{Introduction---}
Terahertz (THz) radiation---{lying at frequencies from 0.1 THz to 70 THz}---has sparked a broad interest recently \cite{tonouchi2007,zhang2017a} due to its key relevance for addressing transition frequencies of vibrational and rotational levels in molecules~\cite{nagai2005}, as well as single-particle and collective transitions in semiconductor materials~\cite{nashima2001}. Such potential provides an avenue to harness light-matter interactions with relevant applications (primarily related to imaging and spectroscopy) in multiple areas, ranging from food sciences \cite{afsah-hejri2019}, medical diagnostics, and biology \cite{woodward2002}, to high-bandwidth communication \cite{hirata2006} or security \cite{kawase2003}.

However, quantum THz technology is at a much more incipient stage than its visible, near-infrared or microwave counterparts \cite{walmsley2015,gu2017,blais2021}. As already demonstrated in these spectral regimes, quantum light offers important technological advantages, such as metrological precision at the Heisenberg-limit \cite{aasi2013enhanced}, alternative quantum computing paradigms \cite{michael2016} or eavesdropping protection in remote communications \cite{gottesman2004security}. Through the development of THz quantum technology, these advances could be transferred and exploited in areas where THz radiation is of key relevance. This avenue would also mean an opportunity to reduce the experimental requirements inherent to current quantum optical implementations, since THz quantum platforms are expected to offer a compromise between the microwave regime, which demands cooling down to millikelvin temperatures and involve important scalability challenges, and the optical one, where materials are strongly absorptive and require nanometric-precision in fabrication. The common mechanism of deterministic single-photon emission enabled by optical dipole transitions in quantum emitters is drastically limited, if not absent, in the THz regime, because the electronic pure dephasing is orders of magnitude larger than the THz emission rate \cite{cole2001}. There are, however, a few demonstrations of heralded quantum THz radiation sources based on spontaneous parametric down-conversion \cite{kitaeva2018}. 

\begin{figure}[t]
\centering
	\includegraphics[width=0.98\linewidth]{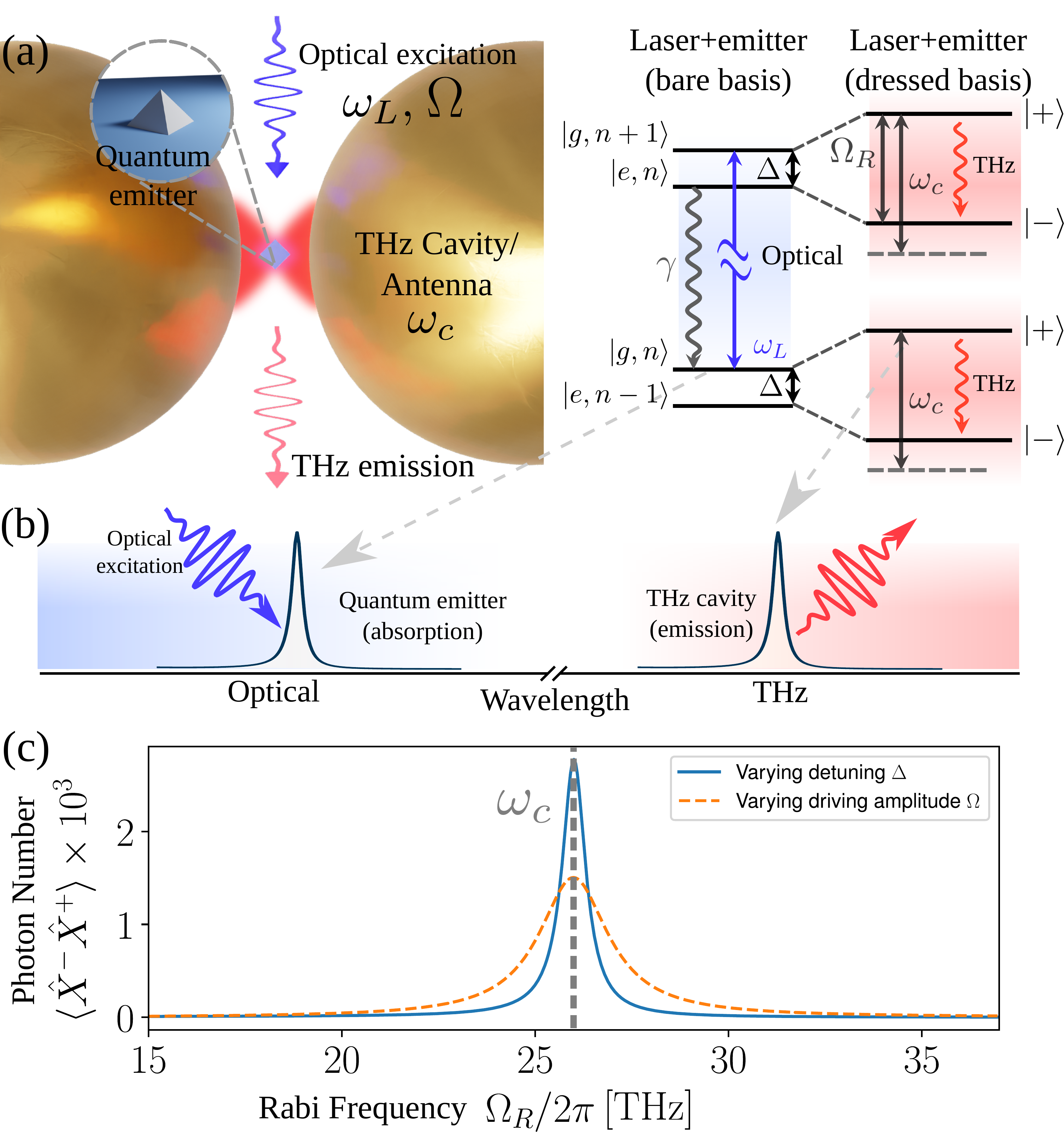}
    \caption{(a) Sketch of a potential experimental implementation with a quantum emitter trapped in a cavity made up by two nanospheres (left) and energy level structure (right): left part represents the bare states basis highlighting energy differences in the optical domain (blue); right side represents the dressed-state basis highlighting THz transitions (red). (b) Absorption and emission properties in the THz and the optical domain. (c) Resonance in the cavity population as the Rabi frequency crosses the cavity frequency for $\{\chi,\kappa,\gamma,\omega_c\}/2\pi=\{0.05,0.158,0.0005,26\}\text{ THz}$. $\Omega_R$ is swept while fixing either the laser amplitude $\Omega/2\pi=10$ THz (blue solid) or the detuning $\Delta/2\pi=10$ THz (orange dashed).}
    \label{fig1}  
\end{figure}

A promising route towards the emission of THz radiation is to exploit the dressing between electronic transitions and driving electric fields, i.e., the AC or dynamical Stark effect. This dressing splits the energy levels into doublets separated by the Rabi frequency $\Omega_R$ [see Fig.~\ref{fig1}(a)], which for certain values of the field intensity can lie in the THz regime. Crucially, in polar systems with broken inversion symmetry, radiative transitions among dressed states in the same Rabi doublet become dipole allowed and have been proposed as a possible channel of emission of THz radiation~\cite{kibis2009,savenko2012,shammah2014,chestnov2017,deliberato2018,pompe2023}. However, to the best of our knowledge, only classical properties of the THz radiation generated---such as the emission spectrum---or semi-classical lasing limits have been considered in such systems. Experimental evidence for such transitions enabled by permanent dipoles exists for Rabi splittings of the order of GHz in superconducting qubits \cite{oelsner2013}.

In this work, we show the prospects of this mechanism with single polar emitters for the realization of quantum optics in the THz regime, demonstrating its ability for the transduction of classical visible light into THz radiation with diverse purely quantum properties, such as single photon emission, multi-photon emission and non-classical correlations between different frequencies of emission. We consider that the single polar emitter is dressed by an optical laser and that its resulting THz transitions ---enabled among the two states of a Rabi doublet--- couple to a THz nanophotonic cavity. The cavity provides a Purcell enhancement of the emission that is eventually radiated into free-space. This design exploits the tunability of the laser parameters and the THz nanocavity architecture to provide considerable brightness and a remarkable optical control of the quantum properties of the emission.

\emph{Model---}
We consider a single two-level system (TLS), consisting of a ground state $|g\rangle$ and an excited state $|e\rangle$, separated by the optical transition frequency $\omega_0$. The TLS is driven by a laser field $\textbf{E}_L$ of frequency $\omega_L$, which is also in the optical range. Furthermore, the TLS couples to a single cavity mode (annihilation operator $\hat a$) with the THz frequency $\omega_c$ and field $\textbf{E}_c$ (cf. Fig. \ref{fig1}(a) for a schematic representation). These features are described by the Hamiltonian ($\hbar=1$): $    \hat H={\omega_0}\hat\sigma_z/2+\omega_c\hat a^\dagger \hat a+\hat{ \textbf{d}}\cdot\textbf{E}_c(\hat a+\hat a^\dagger)+\hat{ \textbf{d}}\cdot\textbf{E}_L\cos(\omega_L t),$
where we have defined the dipole operator $\hat{ \textbf{d}}= \textbf{d}_{ee}(1+\hat\sigma_z)/2+\textbf{d}_{ge}(\hat\sigma_++\hat\sigma_-),$ with
 $\hat\sigma_{\pm,z}$ being the Pauli matrices of the TLS. The term $\propto \textbf{d}_{ee}$ describes the permanent dipole component, originating from asymmetries in the charge distribution of its ground state.

The coherent drive gives rise to two dressed eigenstates of the quantum emitter-laser subsystem, split in energy by the Rabi frequency $\Omega_R=\sqrt{\Delta^2+\Omega^2}$, where $\Delta=\omega_L-\omega_0$ and $\Omega=\textbf{d}_{ge}\cdot\textbf{E}_L$ are the laser detuning and driving amplitude, respectively~\cite{chestnov2017,sanchezmunoz2018b}. These states are given by $|+\rangle = s|e\rangle + c|g\rangle$ and $|-\rangle = -c|e\rangle + s|g\rangle$,  where we define $s=\sin\theta$, $c=\cos\theta$, with $\theta\equiv\arctan(h)\in[0,\frac{\pi}{4}]$, and $h$ is a dressing ratio defined as $h\equiv\frac{\Omega_R-\Delta}{\Omega} \in [0,1]$  that identifies the limit of no dressing ($h=0$) and the resonant limit of a fully-dressed emitter ($h=1$).
The $\hat\sigma_{\pm,z}$ operators can be expressed straightforwardly in terms of the Pauli matrices of the dressed-state basis $\hat\zeta_{\pm,z}$, i.e., $\hat\sigma_{\pm}=cs\hat\zeta_z+s^2\hat\zeta_\pm-c^2\hat\zeta_\mp$ and $\hat\sigma_{z}=(s^2-c^2)\zeta_z-2cs(\hat\zeta_++\zeta_-)$. By applying a rotating wave approximation to eliminate all terms oscillating at optical frequencies in $\hat H$, and then moving to the dressed basis by writing the TLS operators in terms of $\hat\zeta_{\pm,z}$, we obtain the following Hamiltonian~\cite{chestnov2017}
\begin{multline}
\hat H=\frac{\Omega_R}{2}\hat \zeta_z+\omega_c\hat a^\dagger\hat a-2cs\chi(\hat a\hat\zeta_++\hat a^\dagger\hat\zeta_-)\\
    -2cs\chi(\hat a\hat\zeta_-+\hat a^\dagger\hat\zeta_+)+\chi(\hat a+\hat a^\dagger)[1+(s^2-c^2)\hat\zeta_z].
\label{eq:hamiltonian_dressed}
\end{multline}
Here, $\chi=\textbf{d}_{ee}\cdot\textbf{E}_c/2$ is the coupling rate between the TLS and the THz cavity, which, importantly, depends on the permanent component of the dipole moment.
This permanent dipole moment allows for cavity-emitter coupling terms of the form $\propto(\hat a + \hat a^\dagger)\hat\sigma_z$ in the original Hamiltonian, which, crucially, oscillate at THz frequencies, enabling resonant interactions between the THz cavity and the dressed emitter.

Additionally, we take into account cavity photon loss with a rate $\kappa$ and TLS excitation decay with the spontaneous emission rate in vacuum $\gamma$. 
The presence of counter-rotating terms in Eq.~\eqref{eq:hamiltonian_dressed} requires a careful description of the interaction between the system and the bath to prevent unphysical processes such as the emission of photons at zero frequency. In particular, these terms induce a change in the time dependence of the field operator ($\hat a(t)\neq \hat a(0)e^{-i\omega_c t}$), which affects the typical secular approximation commonly made during the derivation of the master equation in the optical regime (similar to the situation found in the ultra-strong coupling 
 regime \cite{beaudoin2011,settineri2018,lednev2023lindblad}). As a result, the interaction between the cavity and the environment is described by the operator
$\hat X^{+}=\sum_{j,k>j}\sqrt{\omega_{kj}/\omega_c}\langle j|(\hat a+\hat  a^\dagger)|k\rangle|j\rangle\langle k|$ that encompasses all the positive-frequency transitions of $(\hat a+\hat a^\dagger)$ \cite{ridolfo2012}. Here, $|k\rangle$ is the $k$-th eigenstate with energy $\omega_k$ (sorted in ascending order) and $\omega_{kj}=\omega_k-\omega_j$. The scaling of $\hat X^+$ with $\omega_{kj}$ is chosen to describe the coupling to an Ohmic bath \cite{settineri2018}. We also define $ [\hat X^{+}]^\dagger=\hat X^{-}$. 
The complete dynamics of the open quantum system is thus described by the Master equation \cite{carmichael2009open} $\dot{\hat\rho}=-i[\hat H,\hat\rho]+\frac{\gamma}{2}\mathcal{D}(\hat\sigma_-)+\frac{\kappa}{2}\mathcal{D}(\hat X^{+})$,
where we have defined the Lindblad superoperator $\mathcal{D}(\hat O)=2\hat O\hat \rho \hat O^\dagger-\hat O^\dagger \hat O\hat \rho-\hat \rho \hat O^\dagger \hat O$, and where the usual decay term $\mathcal D(\hat a)$ has been replaced by $\mathcal D(\hat X^{+})$~\cite{ma2015}. Similarly, the input-output relations are given by $\hat a_\mathrm{out} = \hat a_\mathrm{in} +  \sqrt{\kappa}\hat X^+$~\cite{distefano2018}, so that quantities such as the radiated photon flux will be given by $\kappa \langle \hat X^- X^+\rangle$.
For the case of the emitter, the dressed operator for spontaneous emission remains identical to $\hat\sigma_-$.

In practice, we observe that the standard Lindblad description with $\mathcal{D}(\hat a)$ gives qualitatively the same results as using $\mathcal{D}(\hat X^+)$, given that we are far from being in the ultra-strong-coupling limit $(\chi\ll\omega_c$). On the other hand, the use of the proper input-output relations in terms of $\hat X^\pm$ is crucial, since otherwise one would describe the unphysical emisson of photons with energies equal or close to zero. Even in cases in which these photons only make a minor contribution to the total photon flux emitted, they have a significant impact on the photon statistics, leading to important incorrect contributions to bunched photon statistics when $\Omega_R<\omega_c$.

To gain a better understanding of the dynamics, it is helpful to express the dissipative part in terms of the dressed TLS operators $\hat\zeta_{\pm,z}$. After discarding off-resonant terms based on the assumption that $\omega_c\gg\gamma$, one obtains a  combination of effective incoherent losses, pumping and dephasing, $\dot{\hat\rho}=-i[\hat H,\hat \rho]+\frac{\gamma_-}{2}\mathcal{D}(\hat\zeta_-)+\frac{\gamma_+}{2}\mathcal{D}(\hat\zeta_+)+\frac{\gamma_z}{2}\mathcal{D}(\hat\zeta_z) +\frac{\kappa}{2}\mathcal{D}(\hat X^{+})$, 
 that all depend on the laser detuning, i.e., $\gamma_-=\gamma s^4$, $\gamma_+=\gamma c^4$, and $\gamma_z=\gamma c^2s^2$. It can be shown that this configuration drives the dressed-state population inversion ($\gamma_+>\gamma_-$), if the laser is blue-detuned $\Delta>0$ \cite{chestnov2017}, which is the setting that we will choose for the rest of the paper. In order to achieve a high emission flux, a limit of interest is that of a saturated dressed emitter, reached when the pumping rate  greatly exceeds its decay, $\gamma_+\gg \gamma_-$. This situation takes place when the driving detuning is much larger than the Rabi doublet splitting ($\Delta\gg \Omega$), corresponding to a small dressing ratio $h\ll 1$.

\emph{Resonant mechanism of THz emission---}By tuning the Rabi frequency $\Omega_R$ in resonance with the cavity frequency, $\omega_c$, Jaynes-Cummings-like terms $\propto \hat a\hat\zeta_+ + \hat a^\dagger\hat\zeta_-$ in Eq.~\eqref{eq:hamiltonian_dressed} become resonant and dominate the dynamics. 
In this regime, the system becomes efficient at absorbing optical radiation from the driving field and emitting THz photons, since intra-doublet THz transitions are Purcell-enhanced by the cavity. This regime of operation is sketched in Fig. \ref{fig1}(b) and demonstrated in Fig.~\ref{fig1}(c), which shows the substantial increase in the cavity population when $\Omega_R$ is tuned into this resonant regime. Around this point of operation, we can ignore off-resonant terms in Eq.~\eqref{eq:hamiltonian_dressed} (provided $\omega_c\gg\chi$), and use the resulting effective Jaynes-Cummings Hamiltonian for the dressed states $\hat H={\Omega_R}\hat \zeta_z/2+\omega_c\hat a^\dagger\hat a
    -2cs\chi(\hat a\hat\zeta_++\hat a^\dagger\hat\zeta_-)$. Since under this approximation we have neglected counter-rotating terms, we can safely substitute $\hat X^+$ by $\hat a$ in the Lindblad term of the master equation and in the calculations of photon flux.
    This substitution enables us to obtain approximate analytical solutions, which provide valuable insights into the different emission regimes.

For this analytical calculation, we can assume that the cavity is nearly empty and treat it as a TLS (truncating the number of excitations at 1). Then, we obtain that the photon flux in the resonant condition ($\Omega_R=\omega_c$) is given by:
\begin{equation}
   \kappa \langle \hat a^\dagger \hat a\rangle = \frac{\kappa}{\tilde\kappa}\left(\frac{\gamma_+}{1+\tilde C^{-1} - 4\gamma_z/\tilde\kappa} \right),
   \label{eq:na_analytical}
\end{equation}
where we introduced the effective cooperativity $\tilde C\equiv\frac{16\chi^2}{\kappa\gamma}\left(h^2 + h^{-2} \right)^{-1}$ and an effective rate $\tilde\kappa \equiv \gamma_++\gamma_-+4\gamma_z+\kappa$.
The full expression as a function of $\omega_c$, which can be found in the Supplemental Material (SM)~\footnote{See {Supplemental Material} at [URL will be inserted by publisher] for more information on analytic results, filtered photon statistics, tunability via the laser amplitude, multi-photon resonances, effects of the dressed-state master equation, full electrodynamic simulations, potential experimental setups, minimum {laser} amplitudes, thermal emission and the time scale of the degree of coherence. {It includes Refs}. \cite{10.1063/5.0048049,schoelkopf1999,sclar1984,colautti2020,lange2023b,hadfield2009}.}

,
describes a Lorentzian centered around $\Omega_R$ as shown in Fig. \ref{fig1}(c).

Notice that the introduced effective cooperativity $\tilde C$ is closely connected to the standard expression of the cooperativity, $C=4\chi^2/\kappa\gamma$, but accounts for the effective coupling between the cavity and the dressed emitter, which depends on the detuning between emitter and drive via $h$, so that $\tilde C = 4 C/(h^2+h^{-2})$. In the strongly detuned case $h\ll 1$, we have $\tilde C \approx 4Ch^2$. 
To provide an understanding of the relationship between these quantities, notice that a typical value of $h$ for the parameters chosen in the text is $h\approx0.2$, meaning that $\tilde C \approx 0.16 C$.
A natural limit to consider is when cavity losses represent the dominant decay channel, $\kappa \gg \gamma$, which implies that $\tilde\kappa \approx \kappa$. In that case, in the limit of small cooperativity, $\tilde C \ll 1$, the photon flux acquires the simple form $\kappa\langle\hat a^\dagger\hat a\rangle \approx \gamma_+ \tilde C$, meaning that the flux will increase as $\kappa$ is decreased (so that $\tilde C$ is increased). On the other hand, if the cooperativity is large $\tilde C \gg 1$, we find that $\kappa \langle \hat a^\dagger \hat a\rangle \approx \gamma_+$.
The flux reaches a maximum value when $\kappa$ is decreased into the strong-coupling region $\kappa = 4cs\chi$, an exact value that we obtain by optimizing Eq.~\eqref{eq:na_analytical}.
This maximum flux is, again, simply given by $\kappa\langle  \hat a^\dagger\hat a\rangle\approx \gamma_+$ in the natural situation of $\gamma\ll \chi $ and detuned driving, $h\ll 1$.  The fact that the maximum photon flux is given by $\gamma_+$ implies that the brightness of the THz source scales with the \emph{optical} emission rate into free space. This relationship is noteworthy because the optical emission rate is significantly larger than its THz counterpart, since both scale with the emission frequency as $\omega^3$. A more detailed analytical study of the conditions of maximum flux, including a full expression valid for all regimes, are provided in the SM~\cite{Note1}. 
If $\kappa$ is further decreased to the point in which $\kappa \not\gg \gamma$, the photon flux gets  reduced below its maximum value of $\gamma_+$ since $\tilde\kappa \not\approx \kappa$. We then conclude that the condition of operation that provides the highest possible photon flux of $\gamma_+$ is given by the conditions $\kappa \gg \gamma$ and $\tilde C \gg 1$.

\begin{figure}[t]
        \centering
\includegraphics[width=0.49\textwidth]{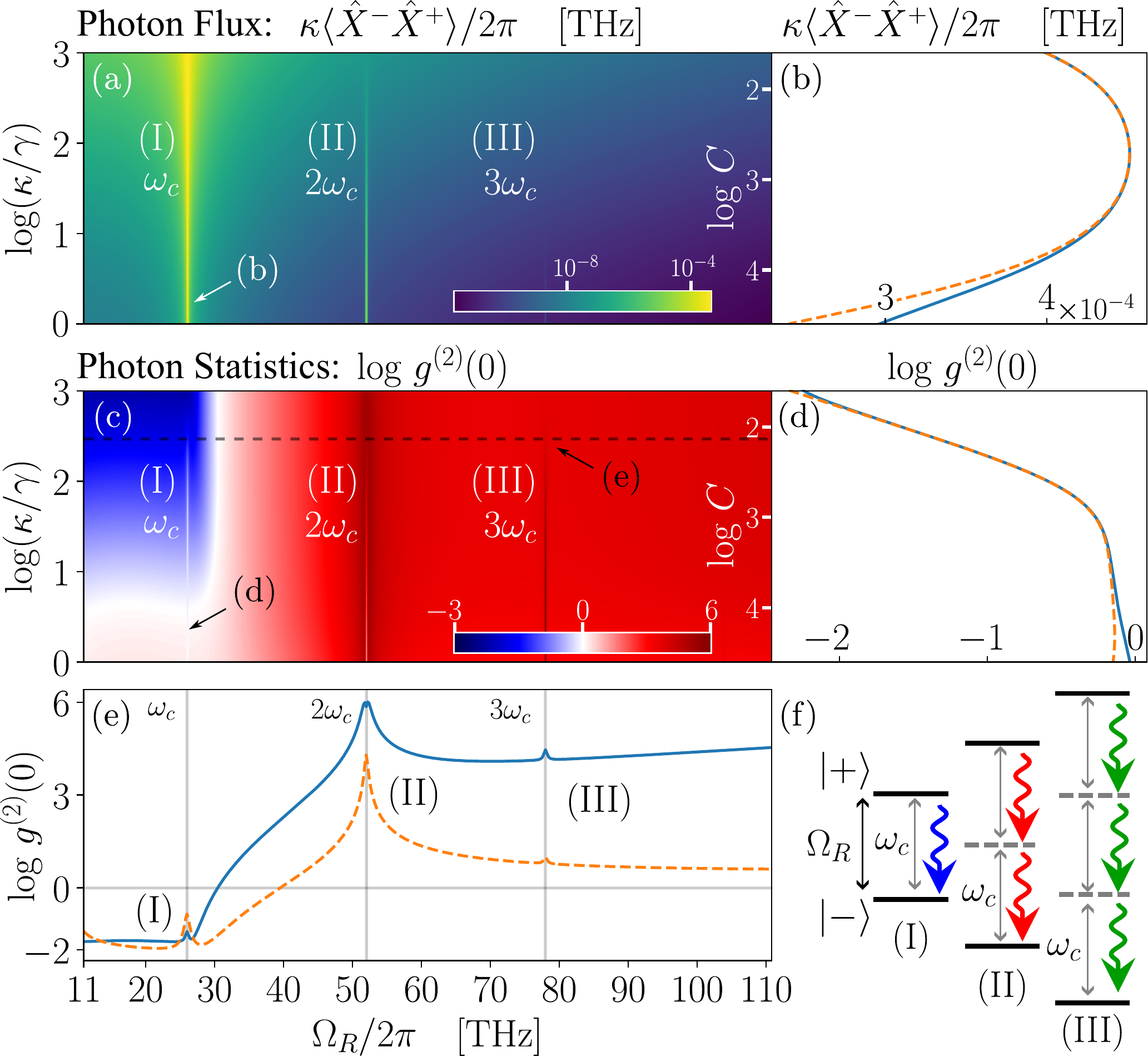}
    \caption{(a) Map of the output photon flux $\kappa\langle \hat X^- \hat X^+\rangle$ as a function of $\Omega_R$ for $\{\chi,\gamma,\omega_c\}/2\pi=\{0.05,0.0005,26\}\text{ THz}$ and fixed $\Omega/2\pi=10$ THz. (b) Shows a cut of (a) for fixed $\Omega_R = \omega_c$, showing both a numerical calculation of $\kappa\langle \hat X^- \hat X^+\rangle$ (blue solid) and the analytic solution for $\kappa\langle\hat a^{\dagger}\hat a\rangle$ (orange dashed) in the Jaynes-Cummings approximation. (c) and (d) reproduce the same map and cut, for the the degree of quantum second-order coherence $g^{(2)}(0)=\langle \hat X^- \hat X^- \hat X^+\hat X^+\rangle/\langle \hat X^- \hat X^+\rangle^2$, with (d) showing results from the simulations (blue solid) and from the analytics (orange dashed). The dashed horizontal line marks the value of $\kappa/\gamma$ chosen for the plot in panel (e).
     (e) $g^{(2)}(0)$ versus $\Omega_R$ with $\textrm{log}(\kappa/\gamma)=2.5$. $\Omega_R$ is changed in two ways: varying $\Delta$ (blue solid), and varying  $\Omega$ with fixed $\Delta/2\pi=10$ THz (orange dashed).  (f) Sketch of the main processes taking place at specific resonances in the maps.}
    \label{fig2}  
        \end{figure}    
These analytical estimations are confirmed by exact, numerical results. 
Fig.~\ref{fig2}(a) shows exact calculations of the output photon flux $\kappa\langle \hat X^- \hat X^+\rangle$ as a function of $\kappa/\gamma$ and $\Omega_R$. On the other hand, Fig.~\ref{fig2}(b) shows the flux at the resonance $\Omega_R=\omega_c$ (labeled I) versus $\kappa/\gamma$ . In both plots, $\Omega_R$ is modified by fixing $\Omega$ and varying the detuning $\Delta$. The orange line in Fig.~\ref{fig1}(b) corresponds to the analytical formula in Eq.~\eqref{eq:na_analytical}, confirming the validity of our analytical results.

Next, we consider the quantum statistics of the emission, measured  through the zero-delay second-order correlation function $g^{(2)}(0)=\langle \hat X^- \hat X^- \hat X^+\hat X^+\rangle/\langle \hat X^- \hat X^+\rangle^2$. We show numerical calculations of its steady-state value in Figs.~\ref{fig2}(c,d).
Notably, we find that the resonance (I) coincides with a regime of strongly antibunched emission where $g^{(2)}(0)<1$, meaning that, in the regime in which the output flux is maximum, \textit{this platform operates as a single THz photon source}. By truncating at 2 excitations, we can obtain an analytic expression for $g^{(2)}(0)$ (see SM~\cite{Note1} for a general expression and further details).
$g^{(2)}(0)$ is antibunched for $\kappa>\gamma$, but when $\kappa$ is decreased into the strong-coupling regime, most of the antibunching will be lost as the system undergoes a lasing phase transition [see kink in the curve in Fig.~\ref{fig2}(d), after which $g^{(2)}(0)$ slowly trends towards 1, i.e., a coherent state]. Note that, at resonance ($\Omega_R=\omega_c$), there is a small region of near-coherent states within the antibunched region, meaning that the antibunching can be made much stronger by setting the cavity slightly out of this resonance.  This effect is more important the lower the $\kappa$, and more visible  in the $\Omega$-ramp in Fig. S1(b) in the SM~\cite{Note1}.

\emph{Multi-photon resonances---}Beyond the main resonant mechanism of THz photon emission at $\Omega_R=\omega_c$ described so far, a sweep over the Rabi frequency as the one shown in Fig.~\ref{fig2}(a,c,e) also unveils additional features in both the output flux and the emission statistics. In particular, one can observe small peaks in the output photon flux when the Rabi frequency $\Omega_R$ is exactly twice (II) or three times (III) the cavity frequency $\omega_c$ (the latter case is barely visible).
 These peaks are related to multi-photon processes enabled by the counter-rotating terms of the form $\hat \zeta_+ \hat a^\dagger$ and $\hat \zeta_z \hat a^\dagger$ in Eq.~\eqref{eq:hamiltonian_dressed}, which we ignored in our analytical derivations presented above.  Each peak corresponds to a $n$-th order process becoming resonant, as has been previously reported in other light-matter systems featuring interaction terms that do not conserve neither parity nor the total number of excitations~\cite{garziano2015,garziano2016,sanchezmunoz2020}. Indeed, at these points, the dynamics are governed by an effective $n$-th order Hamiltonian $\hat H_\textrm{eff}=\lambda_n\left[(\hat a)^n\hat \zeta_++(\hat a^\dagger)^n\hat\zeta_-\right]$, where $n=2\textrm{ or }3$ for (II) and (III), respectively (further information with analytical expressions for $\lambda_n$ can be found in the SM~\cite{Note1}). 
In the presence of dissipation, this gives rise to strongly correlated emission, which in our case corresponds to the simultaneous emission of multiple photons within a Rabi doublet, see Fig.~\ref{fig2}(f). The activation of each of these resonances results in an extraordinary degree of optical tunability of the quantum statistics of the emission, as seen Fig.~\ref{fig2}(e), where, by changing the Rabi frequency of the drive $\Omega_R$, $g_2(0)$ spans eight orders of magnitude from antibunching to superbunching.

The tunability offered when the Rabi frequency $\Omega_R$ is alternatively modified by optically tuning the laser power $\Omega$ instead of its detuning is very similar [cf. dashed orange line in Fig. \ref{fig2}(e)]. However, the limits of $c$ and $s$ are inverted, which leads to bunching for low $\Omega$ and coherent states for large $\Omega$. Further details on the two tuning methods can be found in the SM~\cite{Note1}. Overall, we find that modifying $\Delta$ is a more versatile way to control the system, since the use of strong drivings to reach high values of $\Omega_R$ can result in added pure dephasing (see SM~\cite{Note1}).

\begin{figure}[t]
\centering
\includegraphics[width=0.95\linewidth]{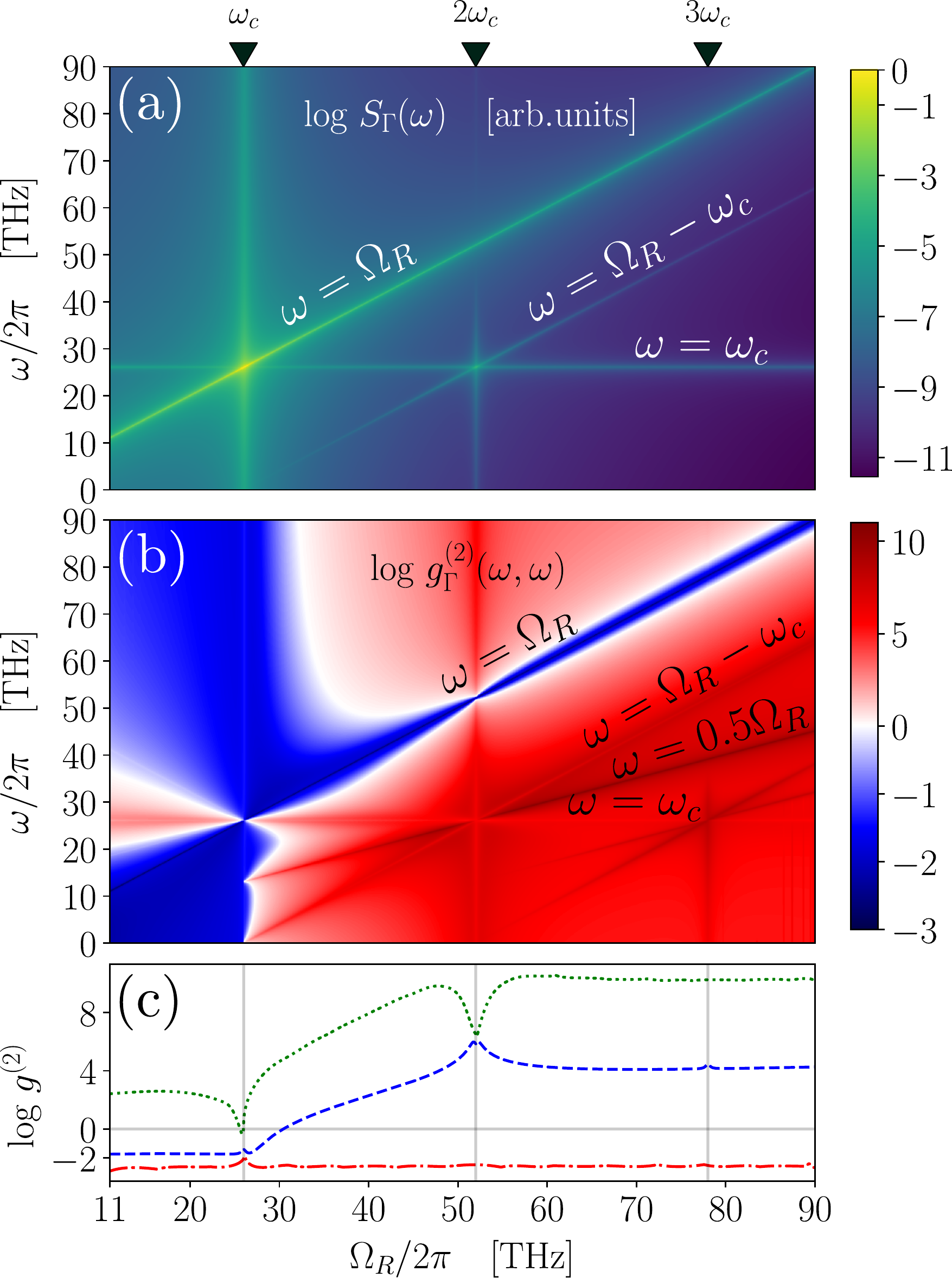}
\caption{Maps of (a) the spectrum $S_\Gamma(\omega)$,  and (b) the degree of quantum second-order coherence $g_\Gamma^{(2)}(\omega,\omega)$ as a function of $\omega$ and $\Omega_R$ for $\{\chi,\kappa,\gamma,\omega_c\}/2\pi=\{0.05,0.158,0.0005,26\}\text{ THz}$. We highlight some of the lines in the maps and denote the corresponding photon frequencies (black). (c) $g^{(2)}(0)$ (blue dashed), $\max_\omega g_\Gamma^{(2)}(\omega,\omega)$ (green dotted) and $\min_\omega g_\Gamma^{(2)}(\omega,\omega)$ (red dash-dotted) for fixed $\Omega/2\pi=10$ THz. }
\label{fig3}
\end{figure}

\emph{Spectral Features---}
Beyond the demonstrated tunability of photon statistics, our proposal can also deliver broadband control over the emission frequency, oftentimes a limiting factor in sources of THz radiation. 
To showcase this feature, we ramp $\Omega_R$ and record the cavity emission spectrum $S_\Gamma(\omega)=\frac{1}{\pi}\int_0^\infty e^{(i\omega-\frac{\Gamma}{2})\tau}\langle\hat X^-(0)\hat X^+(\tau)\rangle d\tau$, where $\Gamma$ is the bandwidth of the sensor, which we take to be equal to $\kappa$.
We focus on a particular case where $\kappa=0.158$ THz, since that value exhibits both strong antibunching and a large output photon flux [see Figs. \ref{fig2}(b) and (d)]. The main frequency of emission is set by the dressed emitter and equal to $\Omega_R$. This feature can be clearly seen in Fig.~\ref{fig3}(a), which shows $S_\Gamma(\omega)$ as the Rabi frequency $\Omega_R$ is varied. This indicates that the Jaynes-Cummings type of dynamics characteristic of the resonance (I) remains important even out of resonance.

A strong secondary signal in the spectrum is observed at the cavity frequency $\omega_c$, regardless of the value of $\Omega_R$. Finally, when $\Omega_R>\omega_c$, a third peak also emerges at a frequency $\omega_2=\Omega_R-\omega_c$, which is a signature of a two-photon processes 
in which the deexcitation of the dressed emitter within a Rabi doublet is accompanied by the emission of a photon at the cavity frequency $\omega_c$ and a second photon of frequency $\omega_2$, matching the energy conservation condition $\omega_c + \omega_2 = \Omega_R$. 
This observation suggests non-trivial dynamics of emission of multi-mode correlated states, which should manifest as strong features the frequency-resolved second-order correlation function at zero delay, $g_\Gamma^{(2)}(\omega_1,\omega_2)$~\cite{delvalle2012a,gonzalez-tudela2013b,ulhaq2012}. To confirm this, we resort to the sensor method develop in Ref.~\cite{delvalle2012a} and  compute this quantity through the correlations between two ancillary qubits, fixing the spectral resolution of these sensors equal to the cavity linewidth $\Gamma=\kappa$ (see SM~\cite{Note1}). We first compute the photon statistics for a given spectral frequency $\omega$, i.e., $g_\Gamma^{(2)}(\omega,\omega)$, versus $\Omega_R$, as shown in Fig. \ref{fig3}(b). We observe that the main emission line $\omega=\Omega_R$ is strongly antibunched, as expected since emission at this frequency stems from first-order processes originating from Jaynes-Cummings-like interaction terms. The other two lines that were clearly visible in the spectrum feature bunched statistics, evidencing their multi-photon character
, and a new strongly bunched line at $\omega=\Omega_R/2$, not visible in the spectrum, is also present. This line corresponds to two-photon processes in which both photons are emitted at the same frequency (instead of one of them being emitted at the cavity frequency). Since this process is not stimulated by the cavity, it is only visible in the statistics. 

These results suggest that frequency filtering can act as an extra control knob of the quantum statistics of the THz emission. Indeed, this is illustrated in Fig. \ref{fig3}(c), where we plot the minimum and maximum possible values of $g_\Gamma^{(2)}(\omega,\omega)$ over $\omega$ for each $\Omega_R$, which ends up always being, respectively, lower or larger than the degree of coherence of unfiltered signal, $g^{(2)}(0)$. The large difference between these maximum and minimum values of $g_\Gamma^{(2)}(\omega,\omega)$ highlights the tunability offered by the method of frequency filtering in the THz regime.

\begin{figure}[t!]
\centering
	\includegraphics[width=0.95\linewidth]{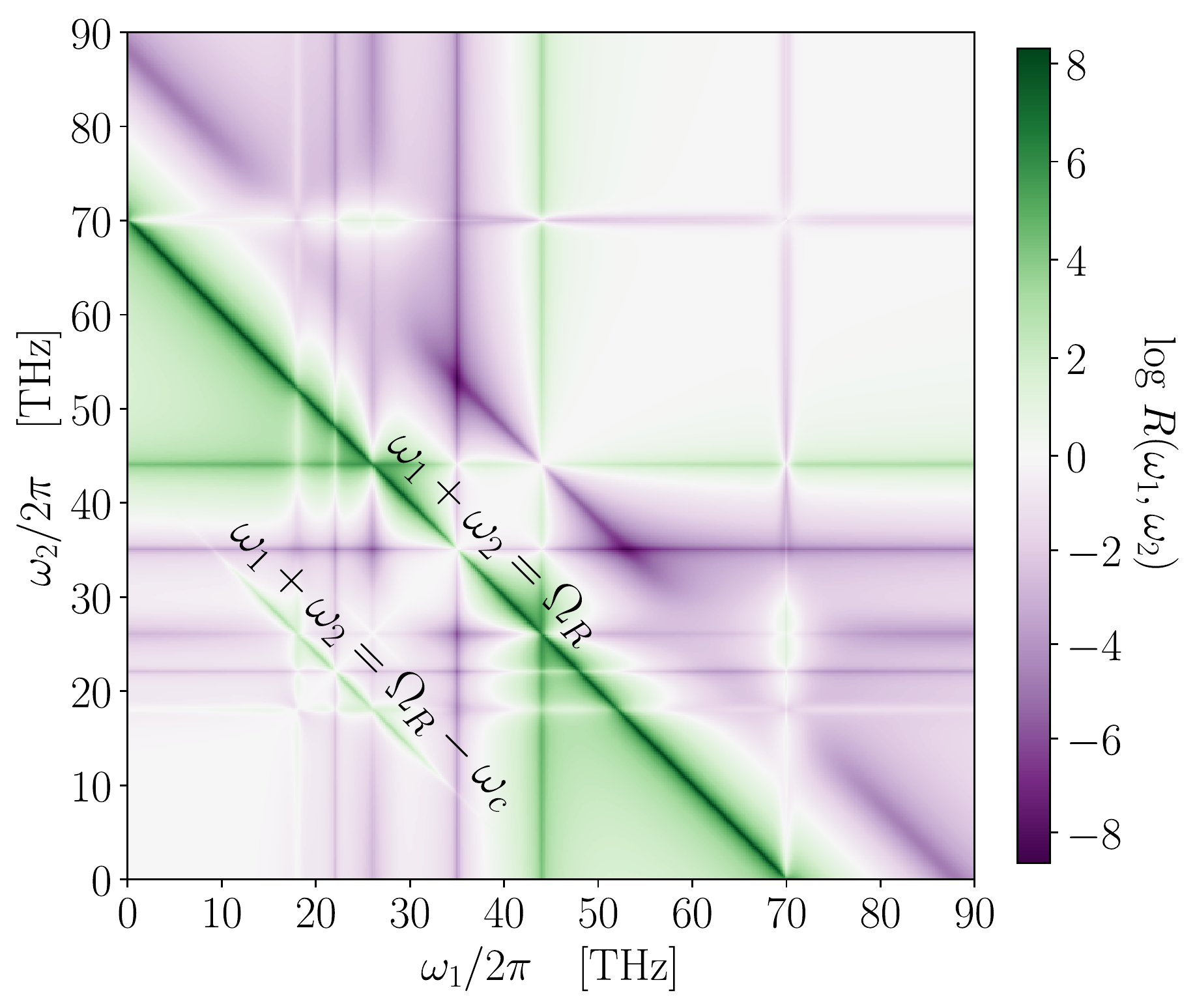}
    \caption{Map of the violation of the
        Cauchy-Schwarz inequality via frequency-resolved correlations. Parameters used: $\{\chi,\kappa,\gamma,\omega_c,\Omega_R,\Omega\}/2\pi=\{0.05,0.158,0.0005,26,70,10\}\text{ THz}$.}
    \label{fig4}  
\end{figure}

Beyond the obvious potential of antibunched THz sources for quantum technologies, spectrally correlated emission like the type we are reporting also holds the potential of quantum applications exploiting non-classical properties such as entanglement~\cite{horodecki2009,kimble2008}. 
To reveal potential non-classical correlations we inspect the cross-correlations between two different frequencies $\omega_1$ and $\omega_2$. Correlations with non-classical character can be identified by the violation of the Cauchy-Schwarz inequality (CSI), reformulated as $ R(\omega_1,\omega_2)=[g_\Gamma^{(2)}(\omega_1,\omega_2)]^2/[g_\Gamma^{(2)}(\omega_1,\omega_1)g_\Gamma^{(2)}(\omega_2,\omega_2))]\leq 1$~\cite{loudon1980,sanchezmunoz2014b,peiris2015}. Fig.~\ref{fig4} shows a typical map of $R(\omega_1,\omega_2)$ in frequency-frequency space, where we chose a relatively large Rabi splitting $\Omega_R/2\pi=70$ THz that allows for multiphoton processes to be observable. 
This map presents a plethora of features that evidences the richness and complexity of the different quantum processes of emission present in this THz source. Providing a complete catalogue of these features is outside of the scope of this text. However, we highlight that the dominant feature exhibiting a strong violation of the CSI is the anti-diagonal line described by the equation $\omega_1+\omega_2=\Omega_R$, corresponding to the joint emission of two photons by the deexcitation of the emitter within a Rabi doublet. For this line one would also find a violation of the Clauser-Horne-Shimony-Holt inequality \cite{clauser1969,sanchezmunoz2014b} (result not shown). In summary, our results suggest that this source can emit entangled THz photon pairs via two-photon processes. Furthermore, we note that our observation of the two-photon resonant peak (II) in the output flux, corresponding to the case $\omega_1 = \omega_2 = \omega_c$, evidences that these processes can be Purcell-enhanced by a cavity in a mechanism akin to previous reports of bundle emission~\cite{sanchezmunoz2014a}.

\emph{Experimental feasibility---}
We now discuss the experimental viability of the single-photon THz sources proposed in this work.
First, we show that the particular set of parameters considered for the calculations in this manuscript, $\{\chi,\kappa\}/2\pi=\{0.05,0.158\}$ THz, 
is readily accessible across a range of platforms. 
The coupling rate $\chi $ is set by the static dipole moment $\textbf{d}_{ee}$ and the electric field strength at the location of the emitter. We consider a value $|\textbf{d}_{ee}|=50$~D as reported in colloidal quantum dots~\cite{doi:10.1063/1.479988}, which are known for their remarkable static dipoles. We have shown via full electrodynamic simulations that, when placed at the $~50$ nm gap between two closely spaced, 1 $\mu$m-diameter spheres \cite{saez-blazquez2022,li2016,zhao2020a} of silicon carbide (SiC) \cite{tiwald1999}, these dipoles provide couplings up to $\chi/2\pi=0.1$ THz, with decay rates $\kappa/2\pi=0.19$ THz (see SM~\cite{Note1}).
Calculations on a nanoparticle-on-mirror geometry \cite{hoang2015,hoang2016} of similar dimensions are also provided, yielding comparable light-matter coupling parameters. These calculations suggest that solid-state emitters with moderate static dipole moments---at least of the order of a few Debyes---can reach interactions strengths comparable to those considered in this work. Such values of static dipole moments have been documented in various systems, including colloidal quantum dots \cite{doi:10.1063/1.479988}, excitonic systems \cite{rapaport2006}, perovskites \cite{lv2021}, simple polar molecules \cite{deiglmayr2010}, macromolecules \cite{kovarskii1999}, non-polar molecules in matrices \cite{moradi2019},  NV centers \cite{Tamarat2006} and Rydberg atoms \cite{doi:10.1126/science.1260722}.

Furthermore, it is worth noting that values different from those considered here could also potentially yield detectable emission of THz radiation. For a fixed cavity configuration, the minimum required value of $\mathbf{d}_{ee}$ to achieve an output power that provides a signal-to-noise ratio of one is a function of the Noise Equivalent Power (NEP) of the detectors used.
We provide the exact relationship in the SM~\cite{Note1}, where we confirm that the output flux provided by the emitters and cavities mentioned above can be detected by a variety of present-day THz detectors.

As a particular example, we can consider current superconducting THz detectors, that can achieve a NEP of up to $10^{-19}$ W Hz$^{-{\frac{1}{2}}}$ with responses below nanoseconds \cite{Sizov_2018}. Together with the bandwidths here considered ($\sim0.16$ THz), these figures yield a minimum detectable power close to $P_\text{min}=\text{NEP}\times\sqrt{\kappa}=4\cdot10^{-14}$ W. Thus, even with a moderate output photon flux of $4\cdot10^{-4}$ THz [cf. Fig. \ref{fig2}(b)], and radiative decays $\kappa^\text{rad}$ of 50\% of the total decay rate ($\kappa=\kappa^\text{rad}+\kappa^\text{abs}$, $\kappa^\text{abs}$ being the absorption rate in SiC), we can estimate an emitted power $P=\kappa^\text{rad}\langle\hat X^-\hat X^+\rangle\hbar\omega_c\approx6\cdot 10^{-13}$ W. 
This estimation amounts to a signal-to-noise ratio of roughly ten, that together with future engineering of emitter interactions on nanostructures and further advances in material science, provide prospects for the creation of bright THz single-photon emitters. Furthermore, since we have shown that the brightness of our source is a function of the linewidth of the emitter, we expect that it could be further amplified via Purcell enhancement by adding a second cavity on resonance with the optical transition of the emitter.

It is also important to consider the feasibility of experimentally measuring photon statistics and establishing the single-photon character of the source we propose. This entails measurements of the second-order correlation function $g^{(2)}(\tau)$, which are typically done via time-tagging in Hanbury-Brown Twiss setups~\cite{somaschi2016a}.
The key figure of merit is the time resolution of the detectors, which need to resolve the intrinsic timescale of the correlations, $\tau_c$. In our source, this timescale is given by $\tau_c \approx \left[(\gamma_- + \gamma_+)(1+\tilde C)\right]^{-1}$ (see Fig. S10 in the SM~\cite{Note1}), which is of the order of tenths of ns for the parameters here considered. Detecting correlations within this timescale can be readily achieved by state-of-the-art THz detectors with ps time resolution and jitter time below 50 ps \cite{loidolt-kruger2021,caselle2014}.

Further potential improvements to all these figures of merit could consist of enhanced nanophotonic architectures, such as hybrid cavities \cite{Gurlek2017} or subwavelength waveguides \cite{martin-cano2010}, as well as the explorations of 2D materials. These can provide THz nanocavities, such as 2D hexagonal boron nitride materials \cite{Autore2018,Caldwell2019}, as well as optical emitting defects \cite{xia2019}.

\emph{Conclusions---} We have shown that a single coherently driven emitter with a permanent dipole moment in a THz cavity can operate as a versatile source of quantum THz radiation, accessing a broad range of frequencies and photon statistics, and featuring a complex quantum correlations between different THz photons.The quantum sources that we propose call for exploring novel interfaces of optomechanical transductions of THz photons to optical ones \cite{roelli2020,xomalis2021,chen2021}, that in conjunction with optical single-photon detectors, or via single electron transistors \cite{komiyama2000}, can open new avenues for the detection of nonclassical THz correlations necessary to harvest the field of THz quantum optics. Beyond the immediate applications of single THz sources for technologies such as imaging or quantum communications, our findings represent a step towards future quantum technologies in the THz, which may consist on more complex cavity setups~\cite{martin-cano2010} capable to enhance the multi-mode correlations that we identify here, and turn them into integrated bright sources of entangled light~\cite{sanchezmunoz2018b,sanchezmunoz2014a} and matter~\cite{gonzalez-tudela2011} at the THz.

\begin{acknowledgments} 
This work makes use of the Quantum Toolbox in Python
(QuTiP) \cite{johansson2012,johansson2013}. We acknowledge financial
support from the Proyecto Sin\'ergico CAM 2020 Y2020/TCS-
6545 (NanoQuCo-CM), and MCINN projects PID2021-126964OB-I00 (QENIGMA) and TED2021-130552B-C21 (ADIQUNANO). A. I. F-D. acknowledges funding from the Europe Research and Innovation Programme under agreement 101070700 (MIRAQLS). C. S. M. and D. M. C. also acknowledge 
the support of a fellowship from la Caixa Foundation (ID 100010434), from the European Union's Horizon 2020 Research and Innovation Programme under the Marie Sklodowska-Curie Grant Agreement No. 847648, with fellowship codes  LCF/BQ/PI20/11760026 and LCF/BQ/PI20/11760018. D. M. C. also acknowledges support from the Ramon y Cajal program (RYC2020-029730-I). 
We thank Vincenzo Macrí for fruitful discussions.
\end{acknowledgments}

\twocolumngrid

\bibliography{citation, THz-references, THz-Quantum-Optics}

\clearpage
\onecolumngrid

  \setcounter{table}{0}  
  \renewcommand{\thetable}{S\arabic{table}} 
  \setcounter{figure}{0} 
  \renewcommand{\thefigure}{S\arabic{figure}}
  \renewcommand{\theequation}{S\arabic{equation}}
  \setcounter{equation}{0}

\begin{center}
{\bf \large Single-photon source over the terahertz regime:\\ Supplemental Material}
\end{center}

\onecolumngrid

\section{Analytic expressions for the Jaynes-Cummings model}
In this section we provide full analytical expressions obtained by solving the master equation with the Jaynes-Cummings Hamiltonian (i.e. with rotating-wave approximations applied), truncated at one cavity excitation.
This truncation is justified by the very small numbers of cavity occupation that we obtain via exact numerical solutions of the master equation.
The full expression for the cavity population that we obtain reads
\begin{equation}
    \langle \hat a^\dagger \hat a\rangle = \frac{16c^2s^2\chi^2\gamma_+ \tilde\kappa}{16c^2s^2\chi^2(\gamma_++\gamma_-+\kappa)\tilde\kappa+\kappa(\gamma_++\gamma_-)[4(\omega_c-\Omega_R)^2+\tilde\kappa^2]},
\end{equation}
which describes a Lorentzian centered around $\Omega_R=\omega_c$.  
At resonance, $\Omega_R = \omega_c$, the expression of the steady-state population of the dressed quantum emitter is given by
\begin{equation}
   \langle\hat \zeta_+\hat\zeta_-\rangle = \frac{\gamma_+[16c^2s^2\chi^2+\kappa(\gamma_++\gamma_-+4\gamma_z+\kappa)]}{16c^2s^2\chi^2(\gamma_++\gamma_-+\kappa)+\kappa(\gamma_++\gamma_-)(\gamma_++\gamma_-+4\gamma_z+\kappa)},
\end{equation}
which for $\gamma\ll\kappa$ simplifies to
\begin{equation}
     \langle\hat \zeta_+\hat\zeta_-\rangle =\frac{\gamma_+}{\gamma_++\gamma_-}\frac{1}{1+\tilde C}=\frac{1}{1+h^4}\frac{1}{1+\tilde C}.
\end{equation}
This exhibits population inversion when $h,\tilde C\ll1$. In the limit $\tilde C\gg 1$, we have    $\langle\hat \zeta_+\hat\zeta_-\rangle\propto \tilde C^{-1}$, which signals a regime in which the dressed-emitter population is depleted via the efficient emission of THz photons through the cavity. 

The exact formula for maximum output flux for the optimum value of $\kappa$ is
\begin{equation}
   \text{max}_\kappa[\kappa\langle  \hat a^\dagger\hat a\rangle]=\frac{16c^2s^2\chi^2\gamma_+}{(4cs\chi+\gamma_++\gamma_-)^2+4(\gamma_++\gamma_-)\gamma_z}.
\end{equation}
In the limit $\chi\gg \frac{\gamma}{4}(c^4+s^4)/sc$, the maximum flux is simply given by $\text{max}_\kappa[\kappa\langle  \hat a^\dagger\hat a\rangle]\approx \gamma_+$.
Finally, a more general expression for the degree of quantum second-order coherence (from a model truncated at two cavity excitations), assuming $\gamma_-,\gamma_z\approx0$, is given by
\begin{equation}
\begin{split}
    g^{(2)}(0)=&2 (\gamma_++2 \kappa ) [\gamma_+ \kappa ^2
 (\gamma_++\kappa )^2 (\gamma_++2 \kappa )
 (\gamma_++3 \kappa )+256c^4s^4\chi^4 (\gamma_+^3+4
\gamma_+^2 \kappa +12 \gamma_+\kappa ^2+6 \kappa
 ^3)\\
  +&16c^2s^2\chi^2 \kappa (\gamma_+^4+5 \gamma_+^3
 \kappa +17 \gamma_+^2 \kappa ^2+23\gamma_+ \kappa
 ^3+6 \kappa ^4)]\\
    /&[\kappa (\gamma_++\kappa )
 (\gamma_++2 \kappa ) (\gamma_++3 \kappa )+32c^2s^2\chi^2(\gamma_+^2+3 \gamma_+ \kappa +3 \kappa^2)]^2.
\end{split}
\end{equation}

\section{Filtered Photon Statistics}
\label{Appendixfilter}
\noindent Here, we elaborate on the numerical method employed for the calculation of frequency-filtered photon statistics.
The calculation is done by coupling the system to two bosonic modes $\hat b_i$ acting as sensors, with energy $\omega_i$ and linewidth $\Gamma=\kappa$. Resorting to the sensor method developed in~\cite{delvalle2012a}, where the cavity mode is extremely weakly coupled ($\epsilon\ll\sqrt{\Gamma\gamma/2}$), $\hat H_c=\epsilon\sum_i(\hat b_i\hat X^-+\hat b_i^\dagger\hat X^+)$, we compute the spectrum and the frequency-resolved degree of quantum second-order coherence by computing expectation values of the sensors, yielding
\begin{equation}
    S_\Gamma(\omega)=\underset{\epsilon\rightarrow0}{\text{lim}}\langle\hat b_1^\dagger\hat b_1\rangle
\end{equation}
and
\begin{equation}
    g_\Gamma^{(2)}(\omega_1,\omega_2)=\underset{\epsilon\rightarrow0}{\text{lim}}\frac{\langle \hat b_1^\dagger\hat b_1\hat b_2^\dagger\hat b_2\rangle}{\langle\hat b_1^\dagger\hat b_1\rangle\langle\hat b_2^\dagger\hat b_2\rangle},
\end{equation}
respectively.

\section{Tunability via the laser amplitude}
\label{AppendixOmegatuning}

\begin{figure}[H]
        \centering
        \includegraphics[width=0.6\linewidth]{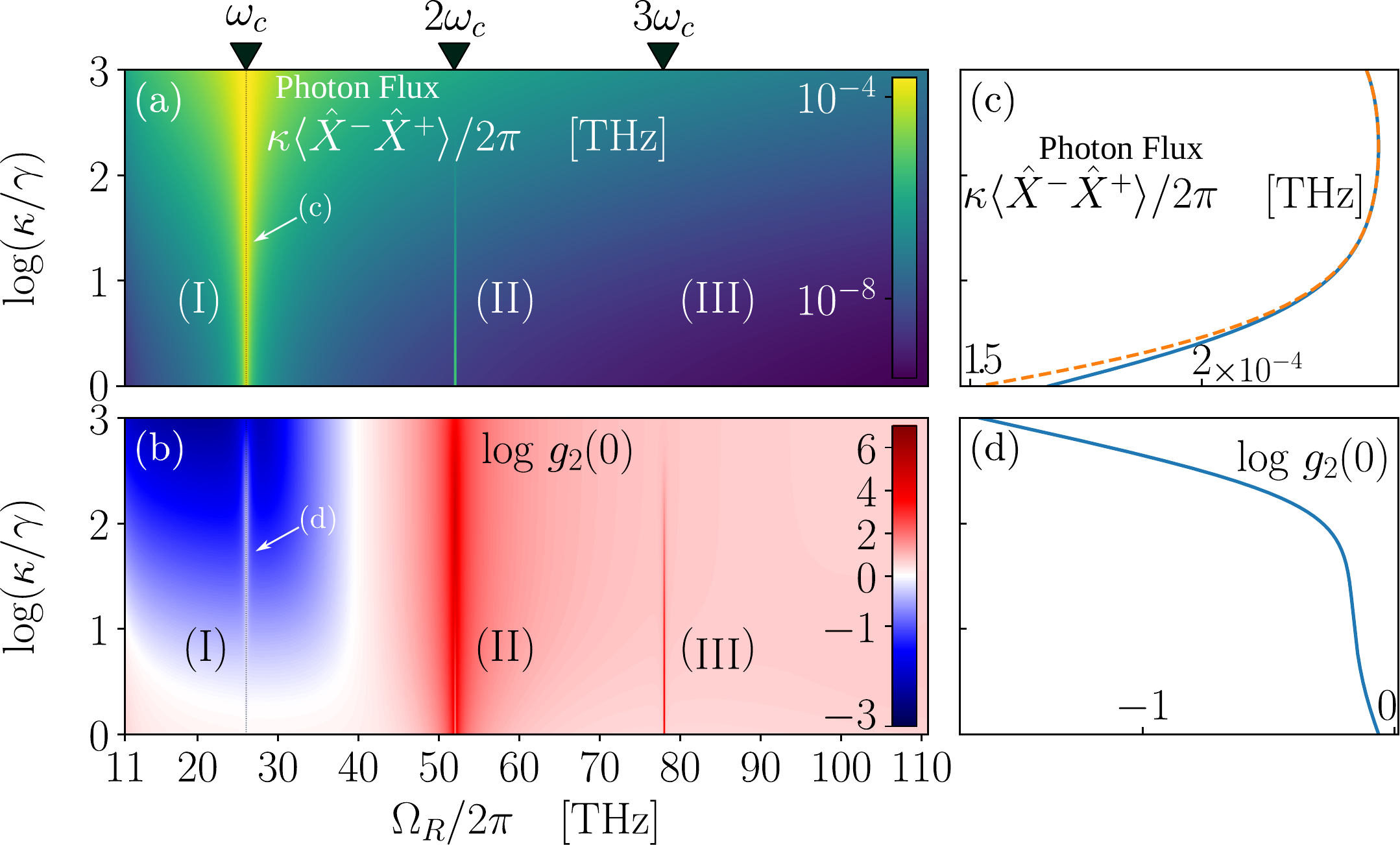}
    \caption{Maps of the output photon flux $\kappa\langle\hat X^-\hat X^+\rangle$ (a) and the degree of quantum second-order coherence $g^{(2)}(0)$ (c) as a function of $\Omega_R$ for $\{\chi,\gamma,\omega_c\}/2\pi=\{0.05,0.0005,26\}\text{ THz}$ for a fixed $\Delta/2\pi=10$ THz. The two adjunct plots (b) and (d) show the scans of the maps along the resonance $\Omega_R=\omega_c$.}
    \label{fig2Omega}
        \end{figure}

\noindent Here we provide further information and results on the implications of modifying the Rabi frequency $\Omega_R$ by tuning the laser amplitude, rather than the laser frequency. Fig. \ref{fig2Omega} is the analogue of Fig. 2 in the main text, except that now $\Delta$ instead of $\Omega$ is kept constant, thus showing an alternative way to tune the quantum statistics with the laser amplitude.
The results are similar, the main differences being an overall lower flux and a lower value of $g^{(2)}(0)$. Notice that, in this situation, $h$ is about three times larger here than in the main text. For instance, in the resonance $(I)$, we obtain $h\approx 0.67$, in contrast to the value $h\approx 0.2$ corresponding to the results presented in the main text.

\section{Validity of the Effective Hamiltonian}
\label{AppendixValidity}
\begin{figure}[H]
        \centering
\includegraphics[width=0.6\textwidth]{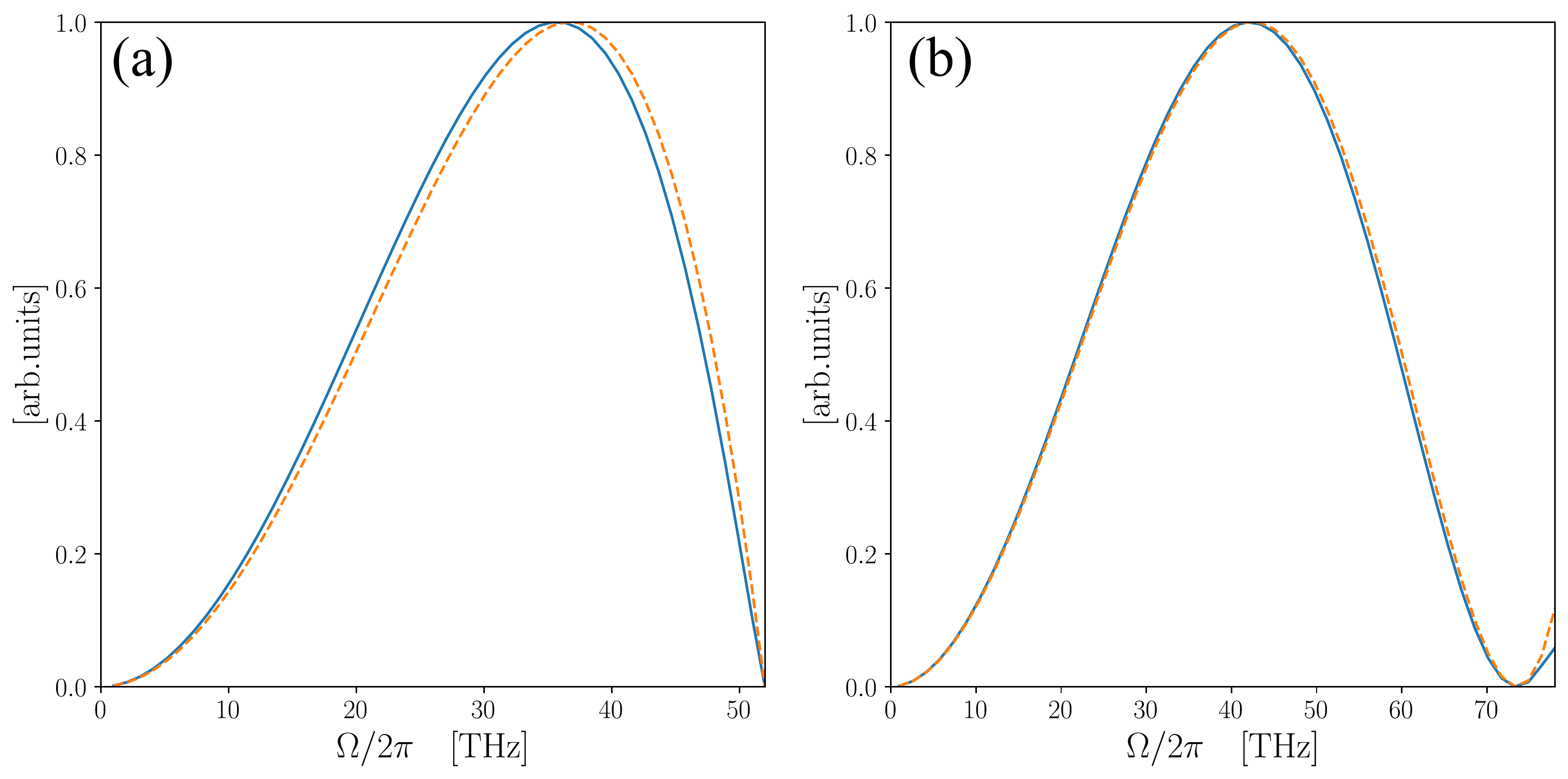}
    \caption{Comparative plots of $|\lambda_n|^2$ (blue solid) and $\langle(\hat X^-)^n(\hat X^+)^n\rangle$ (orange dashed) for $n=2$ (a) and for $n=3$ (b) for varying $\Omega$ and fixed $\Omega_R=n\omega_c$.}
    \label{correlationfunctions}
        \end{figure}

 To check the validity of our assumption that at the specific points the (II) and (III) the Hamiltonian is indeed dominated by terms proportional to $\lambda_n[(\hat a)^n\hat \zeta_++(\hat a^\dagger)^n\hat\zeta_-]$ we compare the ($\Delta,\Omega)$-dependence of the effective coupling strengths $\lambda_n$ obtained via perturbation theory with the $n$-th Glauber correlation function $\langle(\hat X^-)^n(\hat X^+)^n\rangle$, which gives the probability of at least encountering $n$ photons. We have
\begin{equation}
    \lambda_2=\frac{\chi^2}{\omega_c}[cs(s^2-c^2)],
\end{equation}
and
\begin{equation}
    \lambda_3=\frac{\chi^3}{\omega_c^2}[c^3s^3-2cs(s^2-c^2)^2].
\end{equation}
As can be seen in Fig. \ref{correlationfunctions}, there is a good agreement between the effective two- and three-photon transition rates and the correlation functions at second and third order, respectively, validating our interpretation of the results at (II) and (III).

\section{Difference between the Standard and the Dressed Master Equation}

\begin{figure}[tbh]
        \centering
\includegraphics[width=0.49\textwidth]{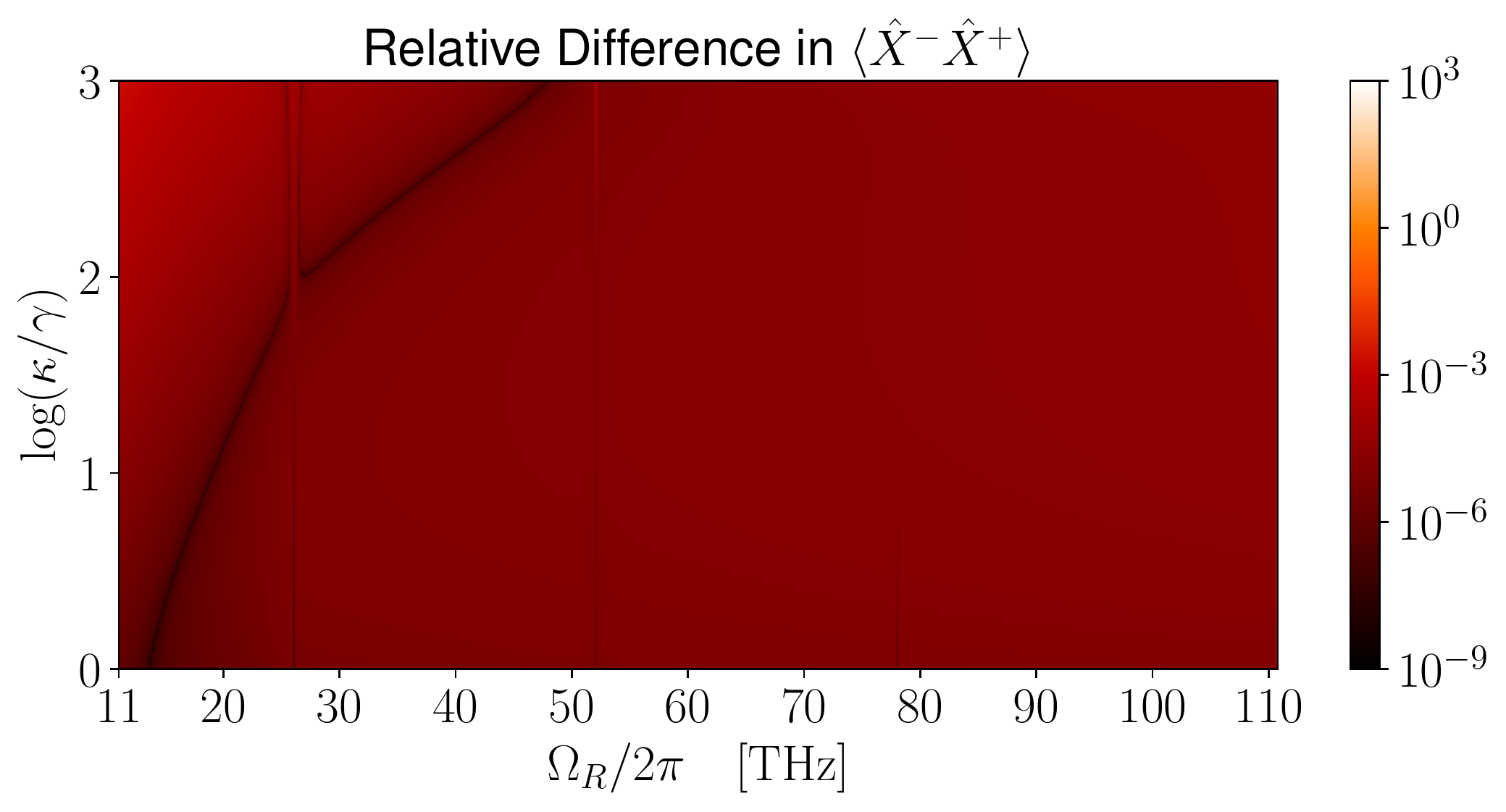}
\includegraphics[width=0.49\textwidth]{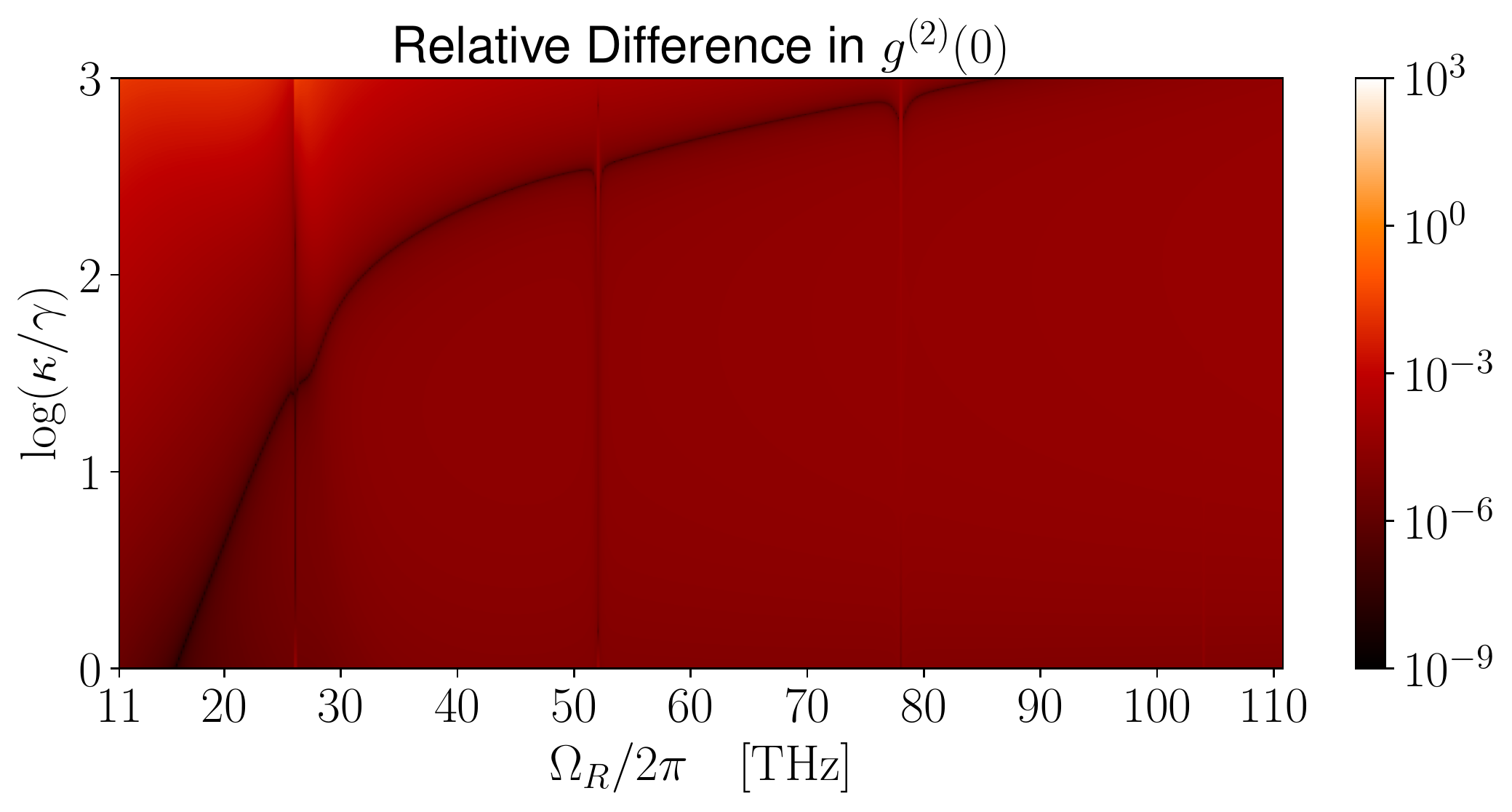}
    \caption{Relative differences in expectation values between simulations using $\mathcal{D}(\hat a)$ and $\mathcal{D}(\hat X^+)$.}
    \label{difflindblad}
        \end{figure}

\begin{figure}[tbh]
        \centering
        \includegraphics[width=0.49\textwidth]{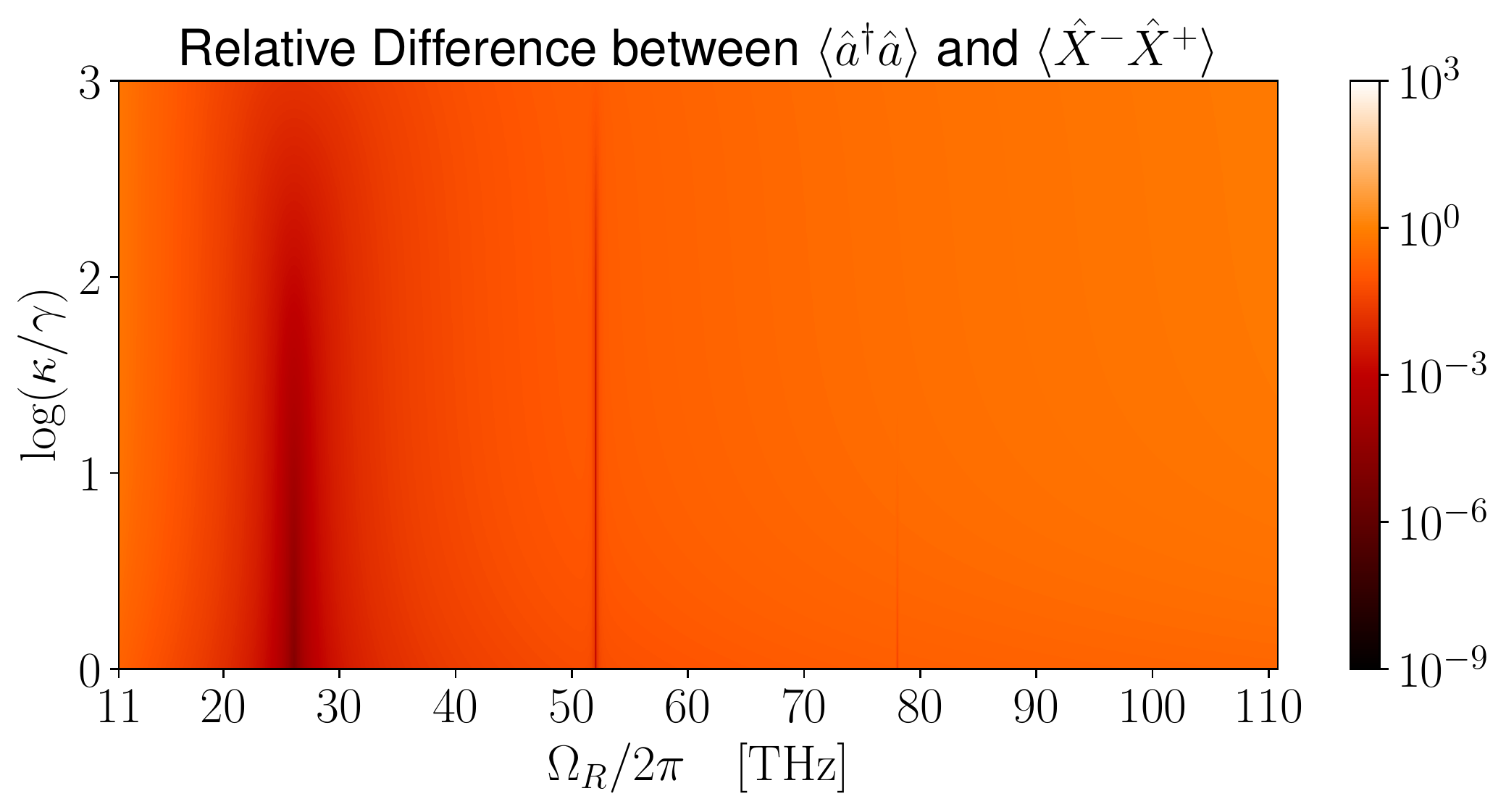}
\includegraphics[width=0.49\textwidth]{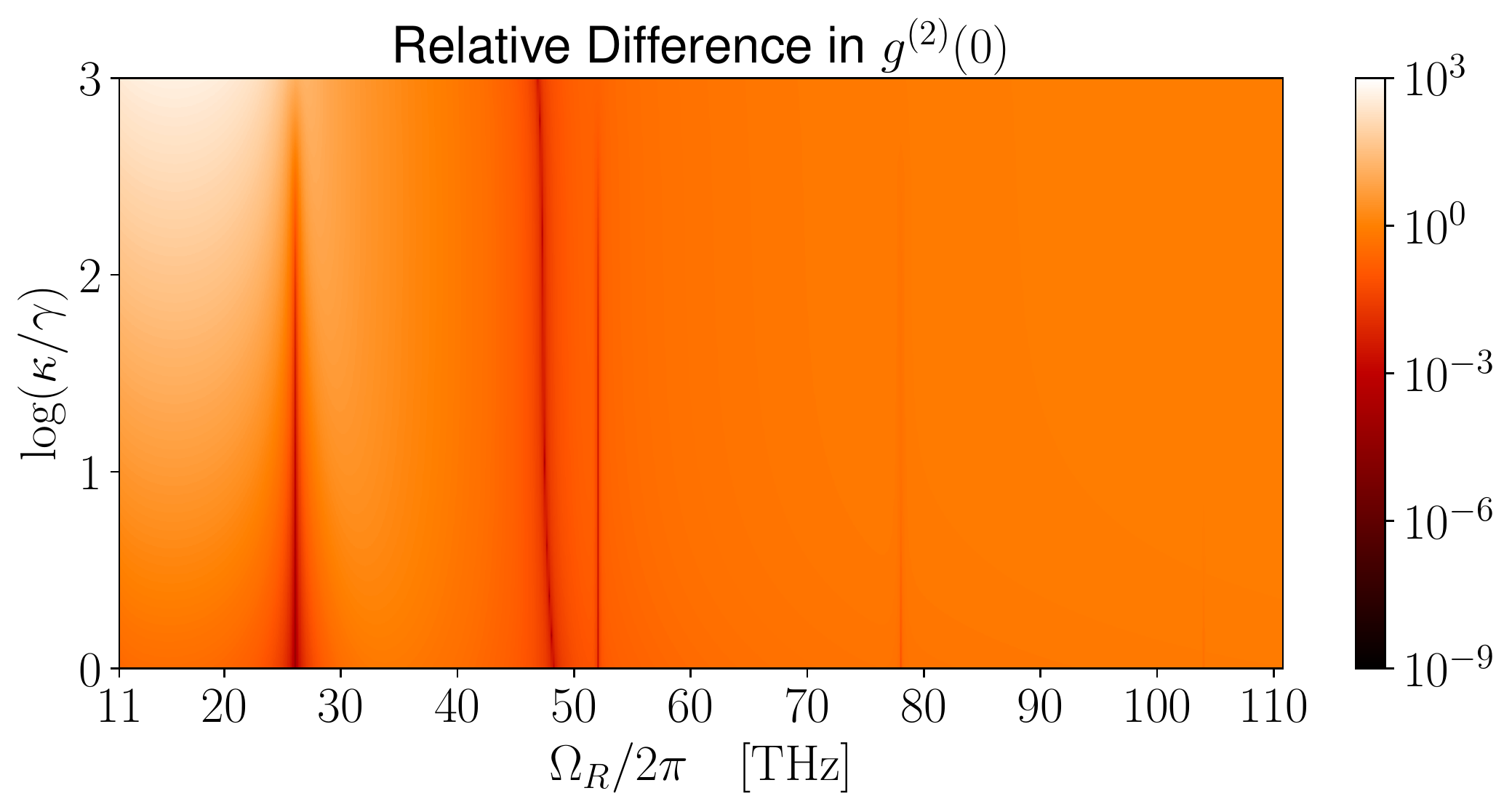}
    \caption{Relative differences between $\langle \hat X^- \hat X^+\rangle$ and $\langle \hat a^\dagger \hat a\rangle$, and between $\langle \hat X^- \hat X^- \hat X^+\hat X^+\rangle/\langle \hat X^- \hat X^+\rangle^2$ and $\langle \hat a^\dagger \hat a^\dagger \hat a\hat a\rangle/\langle \hat a^\dagger \hat a\rangle^2$.}
    \label{diffinputoutput}
        \end{figure}

Fig. \ref{difflindblad} shows that the choice of the Master equation does not have any discernable impact on the results. Fig. \ref{diffinputoutput} shows the major difference that the change in the input-output relation makes. Note how in the resonances (especially $\Omega_R=\omega_c$) the deviation is negligible as long as $\kappa$ is not too large . The largest discrepancy for $g^{(2)}(0)$ is found for $\Omega_R<\omega_c$.

\section{Full electrodynamic simulations of a potential THz cavity}
We propose here a dimer of SiC microspheres as a potential platform for the realization of the THz cavity considered in our model. The frequency-dependent permittivity for this polar crystal, in the vicinity of the Reststrahlen band, can be approximated by a Lorentz oscillator model
\begin{equation}
\epsilon_{\rm SiC}(\omega)=\epsilon_{\infty}+\frac{\epsilon_{\infty}(\omega_{\rm LO}^2-\omega_{\rm TO}^2)}{\omega_{\rm TO}^2-\omega^2-i\omega\Gamma_{\rm SiC}},
\end{equation}
where $\omega_{\rm TO}/2\pi=23.61$ THz and $\omega_{\rm LO}/2\pi=28.91$ THz are the transverse and longitudinal optical phonon frequencies, $\Gamma_{\rm SiC}/2\pi=0.084$ THz is the absorption damping, and $\epsilon_{\infty}=7$ is the static permittivity. These values are taken from the experimental fitting in \cite{tiwald1999}, neglecting anisotropic effects in the SiC response.

\begin{figure}[t]
\centering
	\includegraphics[width=0.6\linewidth]{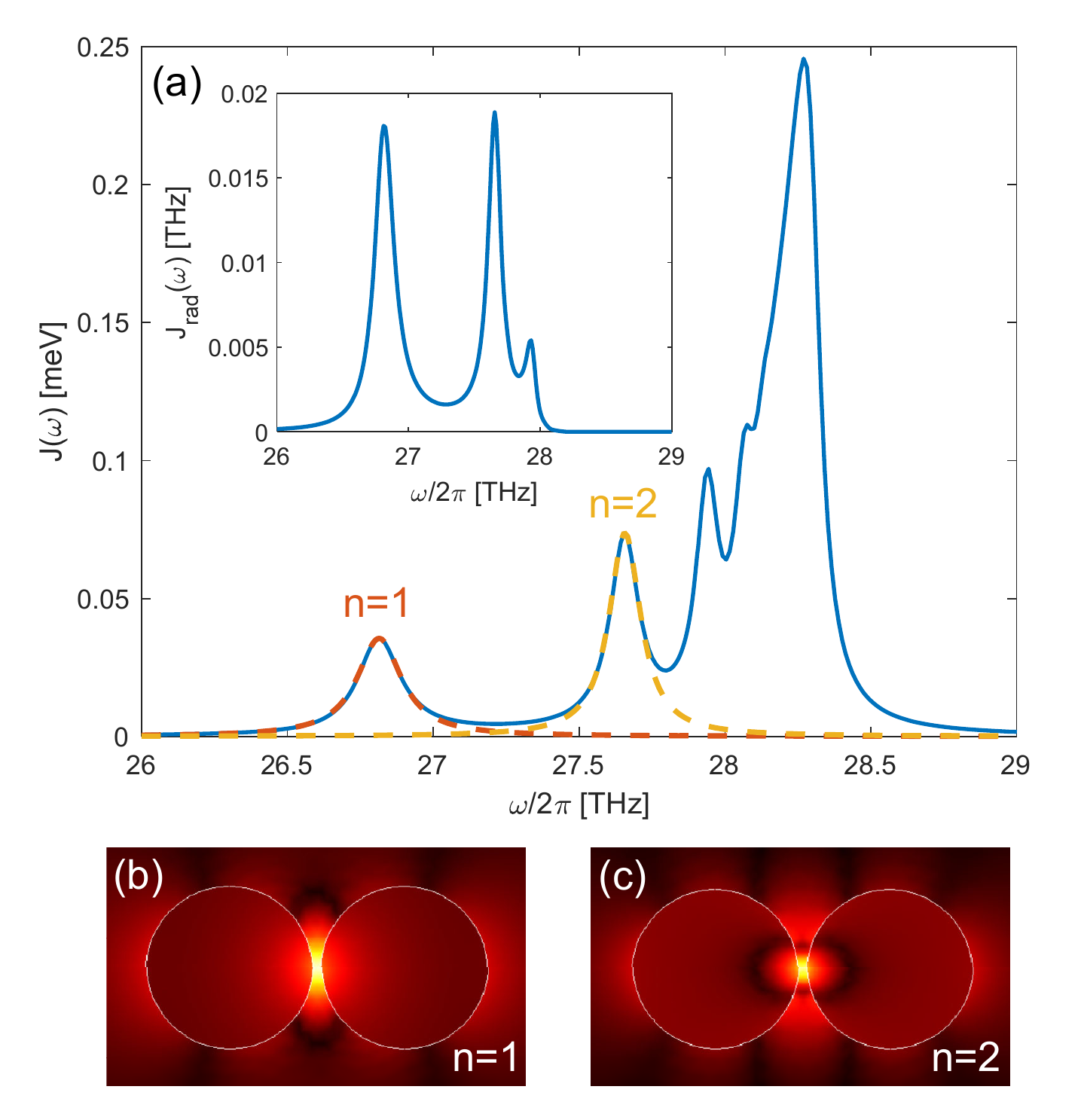}
    \caption{Plot of the spectral density (a), $J(\omega)$ \cite{li2016} at the center of the 50 nm gap between two 1 $\mu$m diameter SiC spheres. The emitter orientation is parallel to the dimer axis and we have taken $|{\bf d}_{\rm ee}|$=50 D for its static dipole moment (also in accordance with experiments \cite{doi:10.1063/1.479988}). $J(\omega)$ was obtained by means of full electrodynamic simulations using the Finite Element Solver of Maxwell's Equations implemented in Comsol Multiphysics. The inset of (a) shows the radiative spectral density for the cavity. Panels (b) and (c) show amplitude maps for the electric field component along the dimer axis and for the $n=1$ and $n=2$ surface phononic modes. Colors render the field amplitude in linear scale from black (minimum) to yellow (maximum).}
    \label{fig5s}  
\end{figure}

The spectral density in Fig. \ref{fig5s}(a) presents a number of peaks, originating from the surface phonon polariton resonances sustained by the cavity. This is defined in terms of the electromagnetic Dyadic Green's function and the static dipole moment as $J(\omega)=\frac{\omega^2}{\pi\epsilon_0\hbar c^2}{\bf d}_{ee}{\rm Im}\{{\bf G(r,r,\omega)}\}{\bf d}_{ee}$. It can be shown \cite{li2016}, that in the quasi-static limit it can be expressed as a sum of Lorentzian terms of the form
\begin{equation}
J(\omega)=\sum_n \frac{\chi_n^2}{\pi}\frac{\kappa_n/2}{(\omega-\omega_n)^2+(\kappa_n/2)^2},
\end{equation}
where $\chi_n$ is the electromagnetic coupling strength for mode $n$, $\omega_n$ its natural frequency, and $\Gamma_n$ its damping rate (including both radiative and absorption channels).

Here, we will only focus on the two lowest-frequency modes ($n=1,2$), which have strong dipolar and quadrupolar characters, respectively. The inset of Fig. \ref{fig5s}(a)
demonstrates that only these two modes contribute significantly to the THz emission from the cavity (weighted by the radiative contribution to the total spectral density~\cite{zhao2020a}). The surface phononic resonances at higher-frequencies, and particularly the pseudomode at 28.25 THz, are dark, and remain effectively decoupled from the far-field of the cavity. Through a Lorentzian fitting of the numerical $J(\omega)$, we can extract the parameters for these two modes:

\begin{table}[h]
\centering
\caption{\bf Parameters of the two contributing modes}
\begin{tabular}{c|c|c|c|c|}
\cline{2-5}
                          & $\omega_n/2\pi$ [THz] & $\kappa_n/2\pi$ [THz] & $\kappa^{\rm rad}_n/2\pi$ [THz] & $\chi_n/2\pi$ [THz] \\ \hline
\multicolumn{1}{|l|}{$n=1$} & 26.815     & 0.186                & 0.101             & 0.102 \\ \hline
\multicolumn{1}{|l|}{$n=2$} & 27.657     & 0.131               & 0.046               & 0.123 \\ \hline
\end{tabular}
\end{table}

Note that we have splitted the mode damping rate, $\kappa_n$, into its radiative, $\kappa^{\rm rad}_n$, and absorption components, and that the latter is given by the loss in the SiC permittivity, $\kappa^{\rm abs}_n=\Gamma_{\rm SiC}$, i.e., $\kappa_n=\kappa^{\rm rad}_n+\Gamma_{\rm SiC}$. Fig. \ref{fig5s}(b) and (c) show maps of the electric field amplitude parallel to the emitter orientation (dimer axis) for the $n=1$ and $n=2$ surface phononic modes, respectively. 

\begin{figure}[H]
    \centering
    \includegraphics[width=0.65\linewidth]{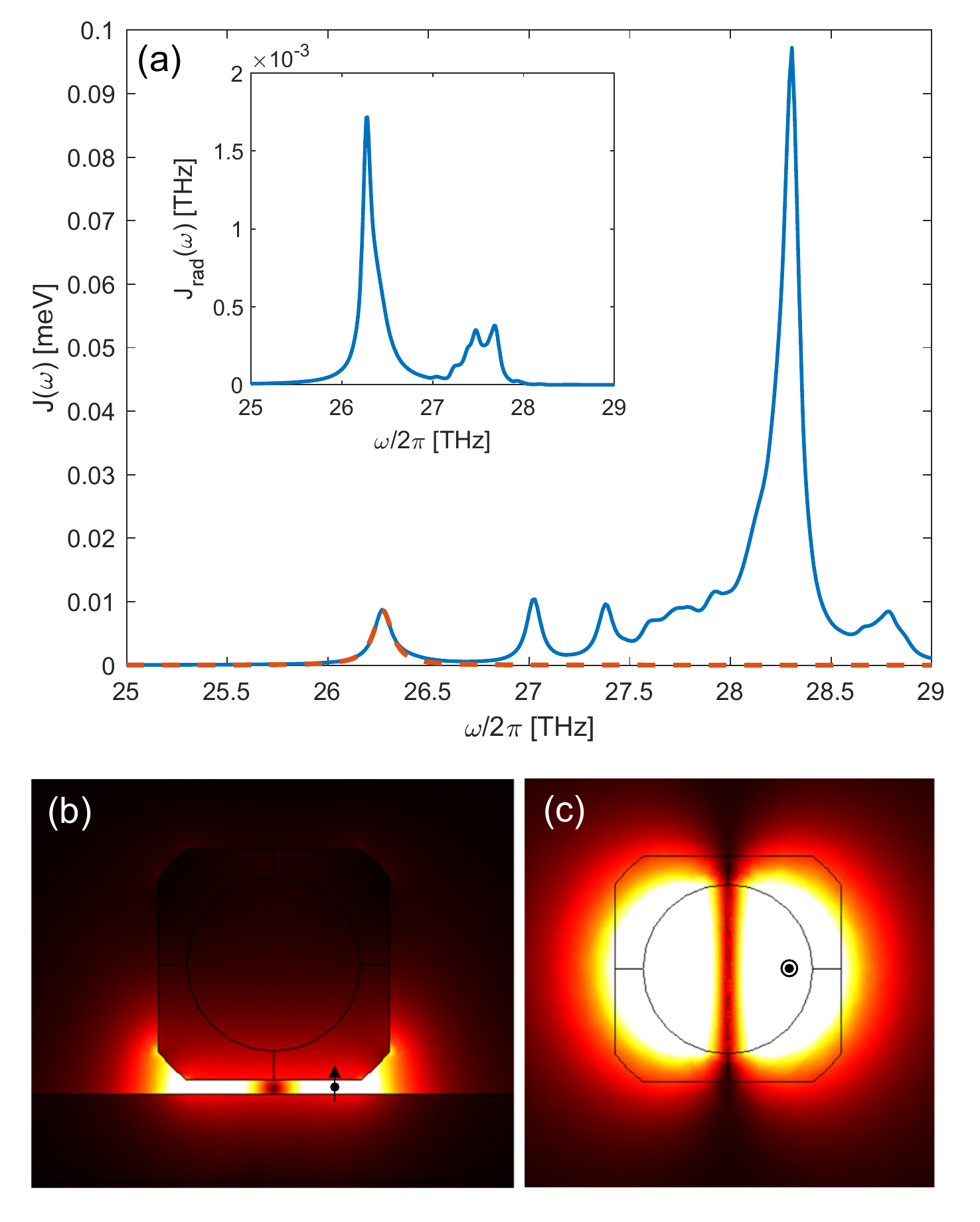}
    \caption{Plot of the spectral density (a), $J(\omega)$ at the 50 nm gap between a 1 $\mu$m side chamfered SiC nanocube and a flat SiC substrate. The emitter orientation is vertical, and its in-plane position is displaced 0.25 $\mu$m from the geometrical center (see black arrows in the panels below). Like in Fig.~\ref{fig5s}, we have taken $|\textbf{d}_{\rm ee}|=50$~D for its static dipole moment. The inset of (a) renders the radiative spectral density for the same configuration, and shows that only the lowest radiative mode contributes significantly to the far-field signal, with an efficiency $\sim22~\%$. Panels (b) and (c) present electric field amplitude maps for this mode in two different cross-sections of the structure. The black dots indicate the emitter position and the black arrow in (b) its orientation. Colors render the field amplitude in linear scale from black (minimum) to yellow (maximum).}
    \label{figS6}
\end{figure}

In recent years, nanocube-on-mirror geometries have attracted attention in the context of plasmonic antennas. Their fabrication is simpler than the nanosphere dimer geometry in Fig. \ref{fig5s}, as they are compatible with chemical deposition techniques, and their planar character makes them suitable for integration with other photonic components. Hoang et al. \cite{hoang2015,hoang2016} have recently shown that their performance for ultrafast light generation can overcome nanosphere dimers. In Fig. \ref{figS6}, we explore this antenna architecture. The system consists of a 1 $\mu$m SiC nanocube with chamfered edges and corners on top of a flat SiC substrate. The gap between them is 50 nm, and the vertically-oriented emitter is placed not at the geometrical center of the gap, but displaced 0.25 $\mu$m, which enables the excitation of a plasmonic mode with a net dipolar moment parallel to the SiC substrate, and therefore directional emission in the vertical direction. Panel (a) plots the spectral density for this geometry and the same static dipole moment as in Fig. \ref{fig5s}. We can observe that the lowest, brightest mode sustained by the geometry is in the same window as the nanosphere dimer, but the inset shows that this lowest energy mode is the only that radiates efficiently out of the structure. Fig. \ref{figS6}(b) and (c) display electric field amplitude maps for this mode within two different cross-sections of the structure. The black dots indicate the emitter position and the black arrow in (b) its orientation. The Lorentzian fitting to $J(\omega)$ for this mode yields $\omega_1/2\pi=26.27$ THz, $\kappa_1/2\pi =0.11$ THz, $\kappa^\text{rad}_1/2\pi =0.028$ THz, and $\chi_1/2\pi =0.040$ THz.

\newpage

\begin{figure}[b]
\centering
	\includegraphics[width=0.8\linewidth]{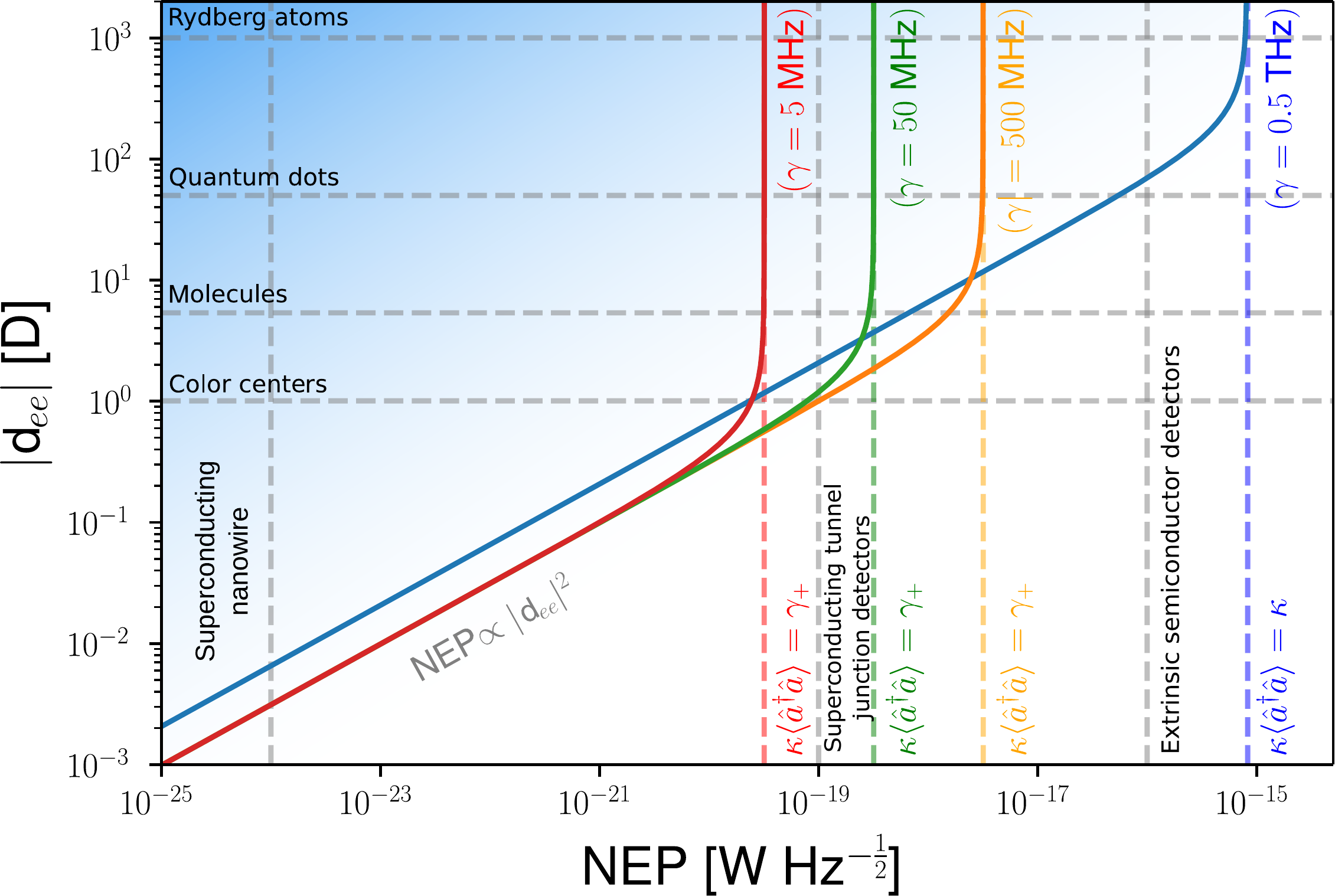}
    \caption{Plot of the critical values of the noise-equivalent power (NEP) and permanent dipole moment, for which the system reaches a SNR of 1 in relation to different detectors/experimental realizations. The different curves correspond to different transition dipole moments/spontaneous emission rates [$\gamma=5$ MHz (red), 50 MHz (green), 500 MHz (orange) and 0.5 THz (blue)]. The horizontal gray lines correspond to (from top to bottom) Cs Rydberg atoms \cite{doi:10.1126/science.1260722}, colloidal semiconductor nanocrystals \cite{doi:10.1063/1.479988}, molecules \cite{deiglmayr2010,moradi2019} and diamond color centers \cite{Tamarat2006}. The vertical gray lines correspond to (from left to right) superconducting nanowires \cite{10.1063/5.0048049}, single-electron devices \cite{schoelkopf1999} and doped Si/Ge detectors \cite{sclar1984}. The NEP for superconducting nanowires was estimated via NEP$=\hbar\omega_c\sqrt{S}$ \cite{roelli2020,hadfield2009}, where $S$ is the dark count rate.}
    \label{fig6s}  
\end{figure}

\section{Potential Experimental Parameters}
Here, we show the relationship between a given detector NEP and the permanent dipole moment which is required to reach a detectable flux, i.e., yielding a signal to noise ratio equal to one. These results were obtained by equating the output power
\begin{equation}
P = \hbar \omega_c \kappa^\text{rad} \langle \hat X^- \hat X^+\rangle    \approx  \hbar\omega_c\frac{\kappa}{\tilde\kappa}(\frac{\gamma_+}{1+\tilde C^{-1}-\frac{4\gamma_z}{\kappa}}),
\end{equation}
with the minimum detectable power
\begin{equation}
P_\text{min}=\text{NEP}\sqrt{\kappa},
\end{equation}
and solving for $|\textbf{d}_{ee}|$---which appears in  $\tilde C$ in the form of the coupling rate $\chi$---essentially demanding a signal-to-noise-ratio (SNR) of 1.
Here, we have assumed that we are in the resonant regime $\Omega_R = \omega_c$ and that $\omega_c \gg \chi$, so that we are in the Jaynes-Cummings regime in which we can substitute $\hat X^+$ by $\hat a$ in the calculation of the flux and we can make use of the analytical equations shown in the main text. Also, for simplicity, we consider here that the decay of the cavity is completely radiative ($\kappa=\kappa^\text{rad}$) and that the detector bandwidth is equal to $\kappa$.

We solve $P/P_\text{min}=1$ for $|\textbf{d}_{ee}|$, while assuming that the permanent dipole can be related with the coupling rate $\chi$ by extrapolating the full electrodynamical simulations, i.e., setting $|\textbf{d}_{ee}|=\chi\cdot [50 \text{ D}]/[0.1\text{ THz}]$. The highlighted spaces above the colored lines are the regions where a certain emitter-detector pair could feasibly produce a measurable outcome. Here we assumed that the emitter frequency is $\omega_0/2\pi=400$ THz and $\kappa/2\pi=0.158$ THz. 
According to Eq.~(2) in the main text, when the permanent dipole is small, the output power is proportional to the cooperativity $\kappa\langle\hat a^\dagger\hat a\rangle=\gamma_+\tilde C$ (or the square of the permanent dipole moment), which corresponds to the quadratic dependence shown at lower values of NEP. When the permanent dipole moment is large, the output power plateaus at $\gamma_+$, and it is thus mainly set by the optical spontaneous emission rate $\gamma$. For very large $|\textbf{d}_{eg}|$, however, $\gamma$ might become larger than $\kappa$, which will result in Eq.~(2) becoming equal to $\kappa\tilde C$ for low $|\textbf{d}_{ee}|$ and $\kappa$ for large $|\textbf{d}_{ee}|$ (see blue line with slightly different behavior). 

Fig. \ref{fig6s} shows that there are several realistic candidates for experimental implementation; for instance, a combination of quantum dots for the emitters and superconducting bolometers for the detection seems like a conservative choice.

\section{Minimum laser amplitudes}

\begin{figure}[hb]
\centering
	\includegraphics[width=0.45\linewidth]{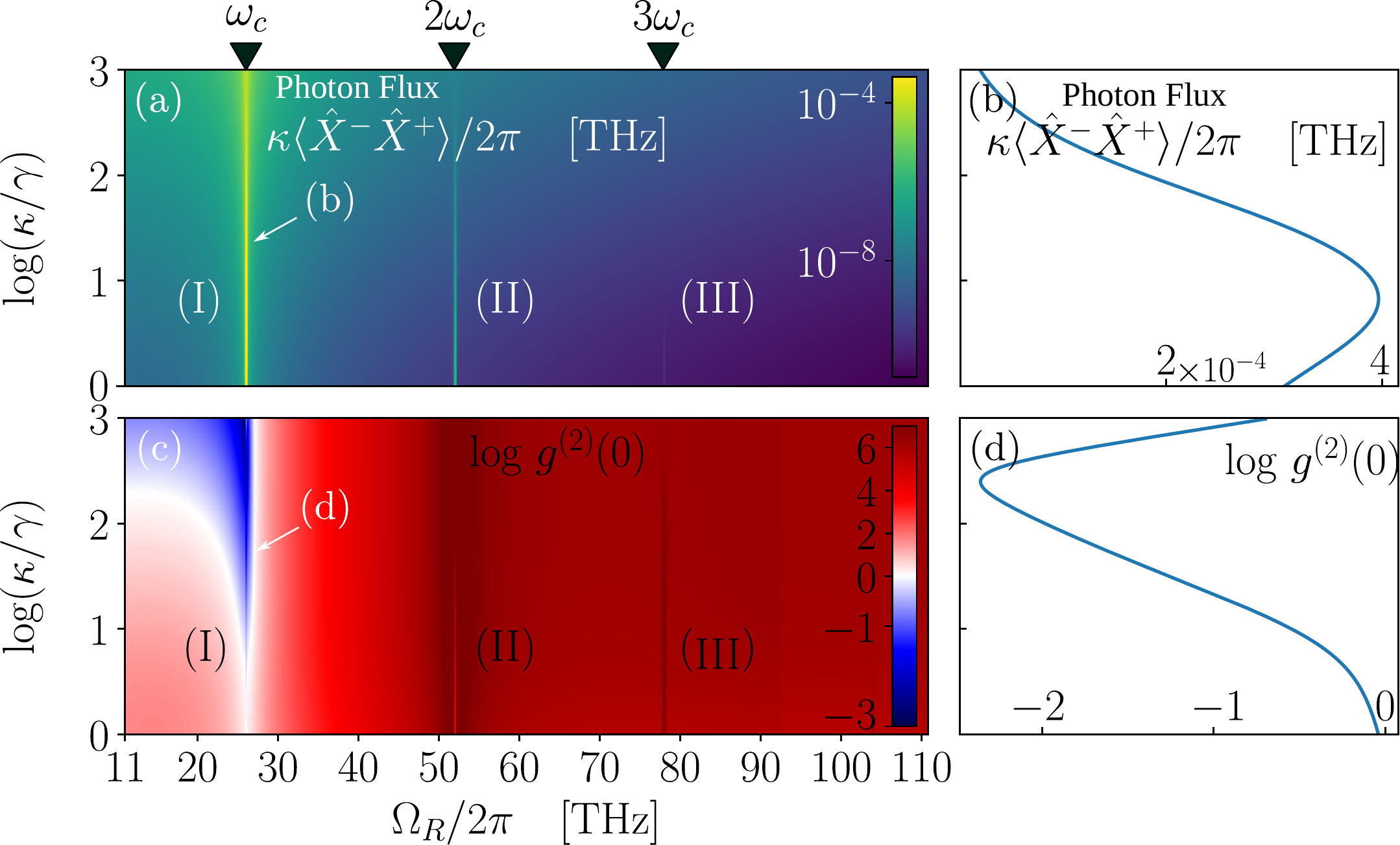}
 \includegraphics[width=0.45\linewidth]{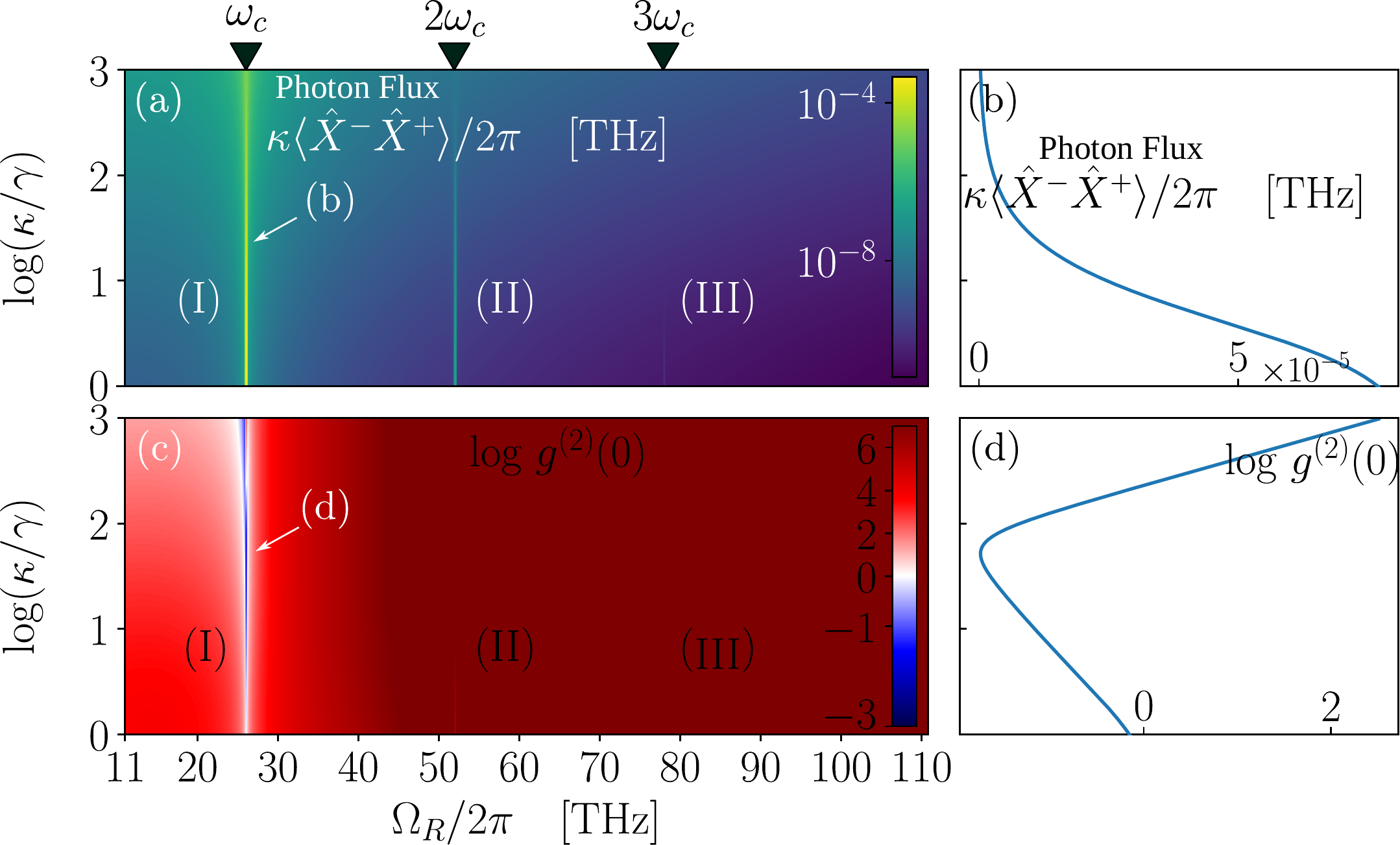}
    \caption{Maps of the output photon flux $\kappa\langle\hat X^-\hat X^+\rangle$ (a) and the degree of quantum second-order coherence $g^{(2)}(0)$ (c) as a function of $\Omega_R$ for $\{\chi,\gamma,\omega_c\}/2\pi=\{0.05,0.0005,26\}\text{ THz}$ for a fixed $\Delta/2\pi=10$ THz, at a temperature of $T=70$ K. The two adjunct plots (b) and (d) show the scans of the maps along the resonance $\Omega_R=\omega_c$. The two figures represent $\Omega/2\pi=1$ THz (left) and $\Omega/2\pi=0.1$ THz (right), respectively.}
    \label{fig7s}  
\end{figure}

Due to the $\theta$-dependence of the interaction, we require that $h\gg1/\sqrt{4C}$ to maximize the flux. This is equivalent to demanding that $\Omega\gg\Omega_R/\sqrt{C}$ (assuming $C\gg1$). The Rabi frequency generated by a Gaussian laser beam is
$\Omega=\sqrt{4\eta P/(\pi w^2)}\frac{|d_{eg}|}{\hbar}$, where $\eta$ is the wave impedance, $P$ the power and $w$ the beam waist. With laser beams with a power of $P=1$ mW focused at $w=390-1500$ nm in SIL (solid immersion lenses) of high NA objectives, as reported in \cite{colautti2020,lange2023b}, Rabi frequencies of the order of $10^3\gamma$ are within reach. Fig. \ref{fig7s} shows the effect of choosing a lower laser coupling $\Omega$. The antibunched region that we had for $\Omega_R<\omega_c$ is pushed back only until it is found in close vicinity to the resonance. In resonance, values of $g^{(2)}(0)$ close to $10^{-2}$ can still be reached. Lowering $\Omega$ too much, however, comes at a cost in brightness and $g^{(2)}(0)$.

\section{Thermal Emission}
\begin{figure}[t]
\centering
	\includegraphics[width=0.6\linewidth]{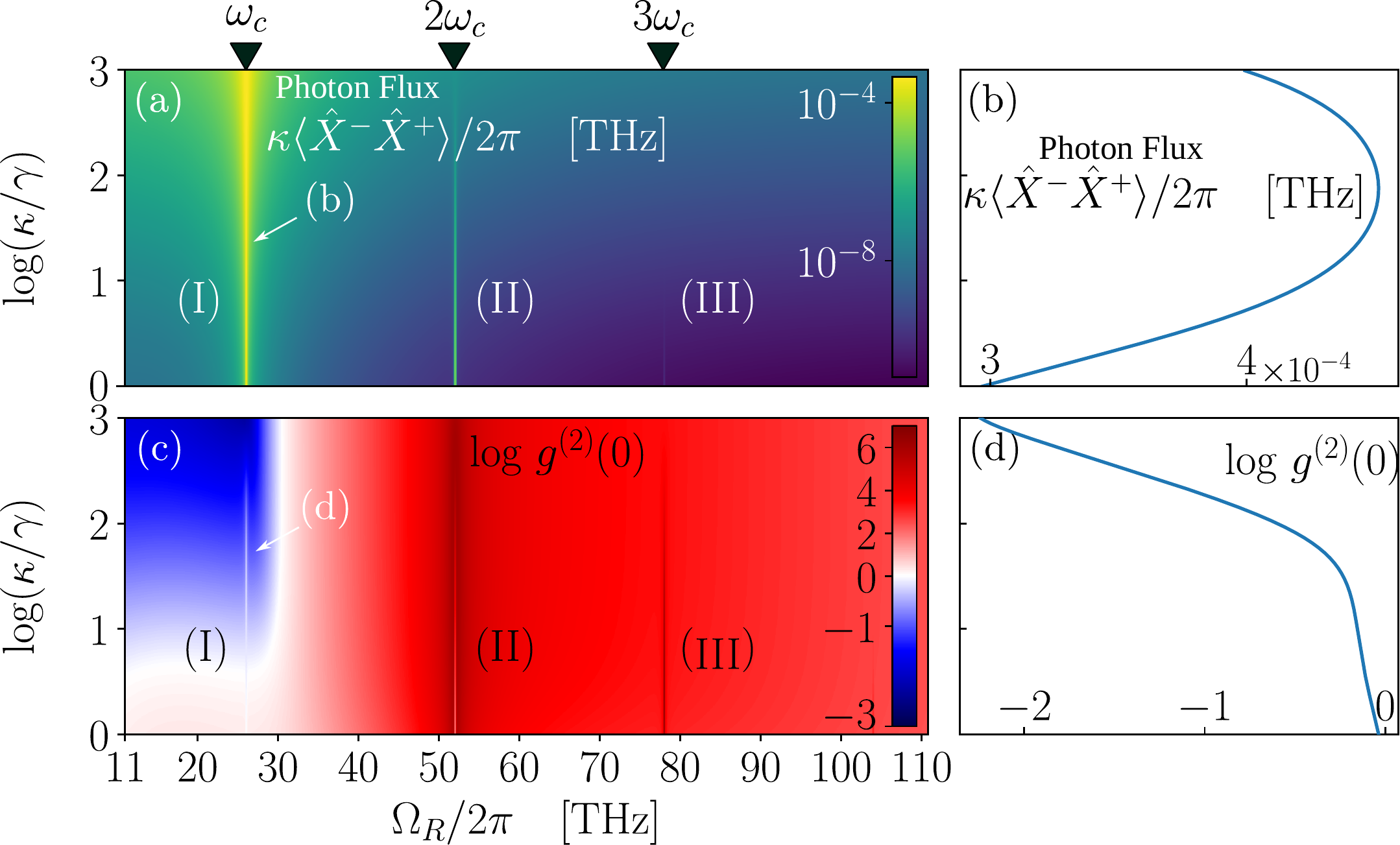}
    \caption{Maps of the output photon flux $\kappa\langle\hat X^-\hat X^+\rangle$ (a) and the degree of quantum second-order coherence $g^{(2)}(0)$ (c) as a function of $\Omega_R$ for $\{\chi,\gamma,\omega_c\}/2\pi=\{0.05,0.0005,26\}\text{ THz}$ for a fixed $\Delta/2\pi=10$ THz, at a temperature of $T=70$ K. The two adjunct plots (b) and (d) show the scans of the maps along the resonance $\Omega_R=\omega_c$.}
    \label{fig8s}  
\end{figure}

We envisage that the system would operate at cryogenic temperatures of $T=70$ K, which corresponds roughly to nitrogen cooling. For that temperature, the thermal occupation number is $\langle n\rangle_{\text{th}}=(e^{\frac{\hbar\omega_c}{k_BT}}-1)^{-1}\approx 10^{-8}$, while the cavity photon number in resonance (the regime that provides antibunching) is roughly $10^{-2}$, meaning that neglecting thermal emission is justified in this regime. Fig. \ref{fig8s} shows a simulation with added thermal emission in the cavity for $T=70$ K with no qualitative difference from the plot at $T=0$ K in the paper (except a decrease in $g^{(2)}(0)$ for large $\Omega_R$). Around $T=200$ K the different features in the map of the quantum second order coherence are almost completely gone and all that remains are thermal states, so room temperature applications are unlikely.

\section{Time-dependent degree of quantum second-order coherence}

\begin{figure}[th]
    \centering
    \includegraphics[width=0.47\linewidth]{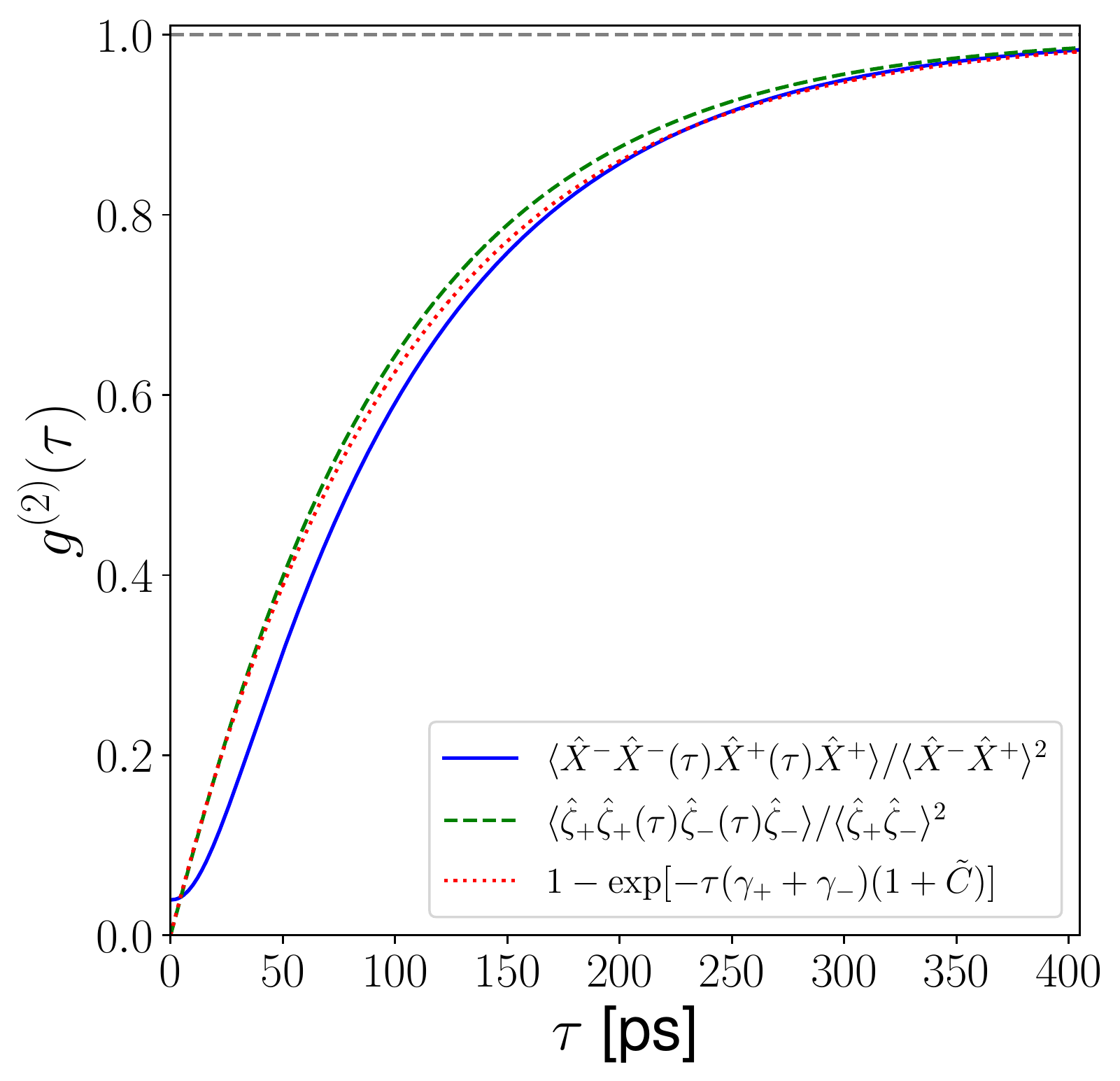}
    \caption{Degree of quantum second-order coherence $g^{(2)}(\tau)$ of the cavity and the emitter as a function of time with an exponential fit. Parameters used: $\{\chi,\kappa,\gamma,\omega_c,\Omega_R,\Omega\}/2\pi=\{0.05,0.158,0.0005,26,26,10\}\text{ THz}$.}
    \label{g2tau}
\end{figure}

When the cavity can be adiabatically eliminated, the dressed emitter is effectively described as a TLS under incoherent pump, with rate $\gamma_+$, and spontaneous emission modified by the cavity, with rate $\gamma_-+(\gamma_-+\gamma_+)\tilde C $. In that case, the second-order correlation function can be obtained analytically and is given by
\begin{equation}
  g^{(2)}(\tau)=1 - \exp\left[-(\gamma_++\gamma_-)(1+\tilde C)t\right].
\end{equation}

This equation establishes a correlation timescale given by 
\begin{equation}
    \tau_c = \left[(\gamma_+ + \gamma_-)(1+\tilde C)\right]^{-1}.
\end{equation}

We have computed numerically the exact values of $g^{(2)}(\tau)$ without any approximations, shown in Fig. \ref{g2tau}. The results, which are reasonably well approximated by the previous equation, confirm that the time delay between subsequent THz emission is given by $\tau_c$, and sets its value to be of the order of 100 ps.

\pagebreak

\end{document}